\documentclass[seceq,twocolumn]{jpsj2} 
\makeatletter

\@addtoreset{equation}{section}
\makeatother

\usepackage{txfonts}
\usepackage{bm}
\usepackage{braket}

\def\runauthor{Hiroaki \textsc{Kusunose}}

\title{
Description of Multipole in $f$-Electron Systems
}

\author{
Hiroaki \textsc{Kusunose}\thanks{kusu@phys.sci.ehime-u.ac.jp}
}

\inst{
Department of Physics, Ehime University, Matsuyama, Ehime 790-8577
}

\abst{
A systematic description of multipole degrees of freedom is discussed on the basis of the Stevens' operator-equivalent technique.
The generalized Stevens' multiplicative factors are derived for all of the electric and the magnetic multipoles relevant to $f$-electron systems.
With extensive use of the Stevens' factors, we express the spatial dependences of the electric and the magnetic fields, and the electric and the magnetic charge densities of localized $f$ electrons.
The latter is utilized to draw wave functions including their magnetic profile in addition to their shape with the charge density.
The definite relation between the operators as quantum-mechanical variables in a multipole exchange model and the multipole moments in expansion of electromagnetic fields is given.
The general treatments for the exchange model with the RPA susceptibility and the Ginzburg-Landau free-energy expansion are discussed, using Ce$_x$La$_{1-x}$B$_6$ as a typical example.
The representative formula of the vector spherical harmonics are summarized, which are suitable basis for vector fields in the spherical expansion.
}

\kword{
multipole, Stevens' multiplicative factor, spherical tensor operator, vector spherical harmonics
}

\begin{document}
\maketitle

\section{Introduction}
Studies with orbital degrees of freedom have encompassed a considerable part of condensed matter physics.
In fields of transition metal oxides a relatively weak spin-orbit coupling provides definite description of spin and orbital, and collective phenomena are investigated in view of an independent or a mutual entanglement of these degrees of freedom\cite{Maekawa04,Kimura03}.
On the other hand, in rare-earth and actinide compounds a strong spin-orbit coupling smears spin and orbital forming a harmonious degrees of freedom called as multipole.

Because of a localized nature of $f$-electron wave function and its large orbital angular momentum, higher multipoles such as octupole become active in orbitally degenerate systems.
They could play a central role for proper understanding on mysterious hidden orders with anomalous responses in thermodynamics and low-energy excitations\cite{icm17}.

It is well known that in macroscopic electromagnetism the concept of multipole is introduced to characterize source charges and currents distributing near the origin\cite{Landau80}.
An atomic-scale counterpart not only characterizes distributions of localized $f$ electrons but also plays a role of a quantum-mechanical variable\cite{Shiina97} as similar to a spin dipole in ordinary magnetism.
Such two aspects of multipole degrees of freedom would be the cause of possible confusion in a practical calculation.
This confusion largely arises from indefinite relation between an expression in terms of the spherical tensor operator\cite{Inui96} and a classical definition of the multipole moment.
Moreover, there exist several different notations with no systematic normalization.
A large number of complicated expressions for higher-rank multipoles may increase occasional errors as well.

The purpose of this paper is to give definite and systematic description of multipoles and clear relation between the quantum-mechanical operators and the classical multipole moments in the expansion of electromagnetic field.
The bridge among these expressions is the Stevens' operator equivalent technique\cite{Stevens52} which is extensively used in an analysis of energy levels under a crystalline electric field (CEF)\cite{Hutchings64}.
In this paper we generalize the so-called Stevens' multiplicative factors to all of electric and magnetic multipoles relevant to $f$-electron systems.
Using the operator equivalent, we provide useful formula for visualization of wave functions including their magnetic profile as well as charge density.
In a realistic situation with or without uniform external fields, we encounter a rather complicated entanglement of plural multipoles.
To analyze such systems systematically, we discuss a general treatment using the random-phase-approximation (RPA) susceptibility and the Ginzburg-Landau (GL) free-energy expansion for a multipole exchange system.
All these would be efficient to explore various phenomena concerning multipole degrees of freedom, and to enhance experimental efforts to quantify measurements using NMR, $\mu$SR, ultrasound, (resonant) X-ray, neutron scattering etc.

The organization of this paper is as follows.
In the next section we demonstrate multipole expansions of scalar and vector potentials\cite{Schwartz55} in which we define the classical multipole moments.
For the given multipole moments we express electric and magnetic fields in a simple form with the vector spherical harmonics\cite{Blatt91}.
The real and the point-group representations are also explained.
In \S3 we discuss the relation between the classical multipole moments and the spherical tensor operators.
The operator equivalent technique based on the Wigner-Eckart theorem bridges over two expressions yielding the generalized Stevens' multiplicative factors\cite{Stevens52,Hutchings64}.
The useful formula for visualization of wave functions are given in \S4.
The fundamental aspect of the multipole exchange system is elucidated on the basis of the RPA susceptibility and the GL free energy in \S5.
In \S6, we illustrate a use of the present arguments with Ce$_{x}$La$_{1-x}$B$_6$ as an example.
The last section summarizes the paper.
There are three appendices.
Appendix A contains the definition and representative formula for the vector spherical harmonics, which are very useful to express vector fields in spherical expansion.
The details of the multipole expansions are given in Appendix B.
The derivation of the generalized Stevens' multiplicative factors is given in Appendix C.
To be self-contained and to provide a coherent notation throughout this paper, we quote several known results with appropriate modifications.

\section{Multipole Expansion}
In this section, we briefly discuss the multipole expansion for the scalar and the vector potentials using the spherical and the vector spherical harmonics as a basis of the expansion.
Through the expansion, we introduce the electric and the magnetic multipole moments, and we examine the symmetry property for them.

\subsection{The scalar potential and the electric multipole moment}
Let us start with the Poisson equation for the scalar potential,
\begin{equation}
{\bm\nabla}^2\phi({\bm r})=-4\pi\rho({\bm r}),
\label{pe_scalar}
\end{equation}
where $\rho({\bm r})$ is the charge density of localized $f$ electrons.
For regions outside the source distribution, the solution of the Poisson equation is expressed as
\begin{equation}
\phi({\bm r})=\sum_{p=0}^\infty \sum_{q=-p}^p \frac{1}{r^{p+1}}Z_{pq}(\hat{\bm r})Q_{pq},
\end{equation}
where we have defined the electric multipole moment as
\begin{equation}
Q_{pq}=\int d{\bm r}\, r^pZ_{pq}^*(\hat{\bm r})\rho({\bm r}).
\label{e_multipole1}
\end{equation}
Here, $\hat{\bm r}={\bm r}/r$ is the unit radial vector, and
\begin{equation}
Z_{pq}(\hat{\bm r})\equiv\sqrt{\frac{4\pi}{2p+1}}Y_{pq}(\hat{\bm r})
\label{norm_sh}
\end{equation}
is the spherical harmonics with the Racah normalization.
Note that we adopt the Condon-Shortley phase, i.e., $[Z_{pq}(\hat{\bm r})]^*=(-1)^qZ_{p-q}(\hat{\bm r})$, yielding that $Z_{p0}(\hat{\bm r})$ of the odd rank $p$ is a real quantity instead of a pure imaginary in the other convention.
The CEF Hamiltonian with the point-charge model is then expressed as
\begin{equation}
{\cal H}_{\rm CEF}=\sum_n q_n\phi({\bm R}_n),
\end{equation}
where $q_n$ and ${\bm R}_n$ represent the charge and the position of the ligand ions, respectively.

The inversion operation transforms $\rho({\bm r})$ to $\rho(-{\bm r})$.
Using $Z_{pq}(-\hat{\bm r})=(-1)^pZ_{pq}(\hat{\bm r})$, we show that $Q_{pq}$ is transformed to $(-1)^pQ_{pq}$.
Namely, the electric multipole moment has the parity $(-1)^p$.
If a system has the inversion symmetry, the odd-rank electric multipole moments vanish since $\rho({\bm r})=\rho(-{\bm r})$ holds.
The time-reversal operation changes nothing on $\rho({\bm r})$.
Therefore, the electric multipole moment is even under time reversal.
Since $\rho({\bm r})$ is real, we have $Q_{pq}^*=(-1)^qQ_{p-q}$.

\subsection{The vector potential and the magnetic multipole moment}
The Poisson equation for the vector potential is given by
\begin{equation}
{\bm\nabla}^2{\bm A}({\bm r})=-\frac{4\pi}{c}{\bm j}({\bm r}),
\label{pe_vector}
\end{equation}
where ${\bm j}({\bm r})$ is the current density originating from the orbital and the spin currents of $f$ electrons.
In contrast to the scalar potential, the vector potential has an intrinsic angular momentum (``spin'') $1$.
Taking this property under consideration, we require basis vector fields which transform like $Z_{pq}(\hat{\bm r})$.
One suitable basis is known as the vector spherical harmonics\cite{Blatt91}.

In the gauge ${\bm\nabla}\cdot{\bm A}=0$, the expansion of the vector potential outside the source distribution has the form
\begin{equation}
{\bm A}({\bm r})=\sum_{p=0}^\infty \sum_{q=-p}^p \frac{1}{r^{p+1}}\left(\frac{{\bm\ell}Z_{pq}(\hat{\bm r})}{ip}\right)M_{pq},
\end{equation}
where the magnetic multipole moment is given by
\begin{align}
M_{pq}=\int d{\bm r}\,{\bm\nabla}\left[ r^{p}Z_{pq}^*(\hat{\bm r}) \right]\cdot{\bm M}({\bm r}).
\label{m_multipole}
\end{align}
Here, ${\bm\ell}=-i{\bm r}\times{\bm\nabla}$ is the (dimensionless) orbital angular momentum, and ${\bm\ell} Z_{pq}(\hat{\bm r})$ is one of the vector spherical harmonics (without normalization) as shown in Appendix A.
${\bm M}({\bm r})$ denotes the magnetization density defined through
\begin{equation}
{\bm j}({\bm r})=c{\bm\nabla}\times {\bm M}({\bm r}).
\end{equation}

By the partial integration in (\ref{m_multipole}), we have
\begin{equation}
M_{pq}
=\int d{\bm r}\,r^pZ_{pq}^*(\hat{\bm r})\rho_{\rm m}({\bm r}),
\label{mm1}
\end{equation}
where we have introduced the magnetic charge density\cite{Blatt91},
\begin{equation}
\rho_{\rm m}({\bm r})=-{\bm\nabla}\cdot{\bm M}({\bm r}).
\end{equation}
This expression is formally similar to that of the electric multipole moment, (\ref{e_multipole1}).

In contrast to the charge density, the magnetic charge density as well as $M_{pq}$ is odd under time reversal.
The inversion operation transforms ${\bm M}({\bm r})$ to ${\bm M}(-{\bm r})$, and consequently $\rho_{\rm m}({\bm r})\to-\rho_{\rm m}(-{\bm r})$.
Thus, the magnetic multipole moment has the parity $(-1)^{p+1}$.
In the presence of the inversion symmetry, the even-rank magnetic multipole moments also vanish since
${\bm M}({\bm r})={\bm M}(-{\bm r})$ [$\rho_{\rm m}({\bm r})=-\rho_{\rm m}(-{\bm r})$].
The complex conjugation is $M_{pq}^*=(-1)^qM_{p-q}$.

\subsection{The electric and magnetic fields}
From (\ref{grad_sh2}), we have the simple expression of the electric field in the multipole expansion,
\begin{equation}
{\bm E}({\bm r})=-{\bm\nabla}\phi({\bm r})=-\sum_{p=0}^\infty\sum_{q=-p}^p \frac{\sqrt{4\pi(p+1)}}{r^{p+2}}Q_{pq}{\bm Y}_{pq}^{p+1}(\hat{\bm r}).
\label{ele_field}
\end{equation}
Similarly, we obtain the expression of the magnetic field,
\begin{equation}
{\bm B}({\bm r})={\bm\nabla}\times{\bm A}({\bm r})=-\sum_{p=0}^\infty\sum_{q=-p}^p \frac{\sqrt{4\pi(p+1)}}{r^{p+2}}M_{pq}{\bm Y}_{pq}^{p+1}(\hat{\bm r}),
\label{mag_field}
\end{equation}
which is formally similar to ${\bm E}({\bm r})$.
For the given electric and magnetic multipole moments, the electric and the magnetic fields are calculated straightforwardly by using the definition of the vector spherical harmonics, (\ref{def_vsh}).
It is also useful to give the scalar and the vector products with $\hat{\bm r}$,
\begin{align}
&
\hat{\bm r}\cdot{\bm E}({\bm r})=\sum_{p=0}^\infty\sum_{q=-p}^p(p+1)Q_{pq}\frac{Z_{pq}(\hat{\bm r})}{r^{p+2}},
\\
&
\hat{\bm r}\times{\bm E}({\bm r})=\sum_{p=0}^\infty\sum_{q=-p}^pQ_{pq}\frac{{\bm\ell}Z_{pq}(\hat{\bm r})}{ir^{p+2}}.
\end{align}
Similar expressions for ${\bm B}({\bm r})$ are obtained by replacing $Q_{pq}$ with $M_{pq}$.

\subsection{The real and the point-group representations}
The multipole with $q\ne0$ in the spherical representation is a complex quantity.
It is useful to introduce the real representation for $q>0$ as follows:
\begin{equation}
\begin{split}
A_{pq}^{\rm (c)}=
\frac{(-1)^q}{\sqrt{2}}\left( A_{pq}+A_{pq}^* \right),
\\
A_{pq}^{\rm (s)}=
\frac{(-1)^q}{\sqrt{2}i}\left( A_{pq}-A_{pq}^* \right),
\end{split}
\label{real_rep}
\end{equation}
where $A_{pq}$ represents any quantity that transforms like $Z_{pq}(\hat{\bm r})$ under spatial rotation.
The corresponding real expressions for $r^pZ_{pq}(\hat{\bm r})$ (called as the tesseral harmonics\cite{Hutchings64}) are summarized in Table \ref{scaled_sh}.
In the real representation, a sum of products is rewritten as
\begin{equation}
\sum_{q=-p}^pA_{pq}^*B_{pq}=
\sum_{q=1}^p\left[
A_{pq}^{\rm (c)}B_{pq}^{\rm (c)}+
A_{pq}^{\rm (s)}B_{pq}^{\rm (s)}
\right]
+
A_{p0}B_{p0}.
\end{equation}
A similar transformation is applicable to vector fields as well.

In reality, magnetic ions are placed in a crystal with a proper point-group symmetry.
When a CEF splitting is small and a total-angular momentum ($J$) multiplet can be treated as a whole, the spherical or the real representation is appropriate.
On the other hand, when a CEF splitting is large and one of CEF multiplets dominates low-energy physics, the point-group irreducible representation is suitable to classify the multipole moments.
Due to the fact that any point group is a subgroup of the rotation group, the point-group harmonics are constructed as linear combinations of the spherical harmonics.
For instance, the cubic harmonics\cite{Lage47,Sugano70} under $O_h$ are given in Table \ref{cubic_sh}, where $\Gamma$ and $\gamma$ represent the irreducible representation and its component in the Bethe notation\cite{Inui96}.
Note that any scalar product in the spherical expansion can be replaced by the point-group identity representation, which has the form with a sum of pairs of the same irreducible representation,
\begin{equation}
\sum_{q=-p}^pA_{pq}^*B_{pq}
\to
\sum_{\Gamma\gamma}A_{p\Gamma\gamma}^*B_{p\Gamma\gamma}.
\label{rot_point}
\end{equation}
In the following sections, we often give results only in the spherical representation.
Any scalar product may be replaced properly by the corresponding point-group representation.

\section{Multipole and Stevens' Operators}
\subsection{Multipole operators}
In the previous section, we have introduced the electric and the magnetic multipole moments, which are determined by the charge density $\rho({\bm r})$ and the current density ${\bm j}({\bm r})$ [or equivalently the magnetization density ${\bm M}({\bm r})$] of the localized $f$ electrons.
In the quantum statistical mechanics, the charge and the current densities should be regarded as a thermal average over $f$-electron states.

The corresponding one-body charge density operator acting on $f$-electron wave functions is given by
\begin{align}
\hat{\rho}({\bm r})=-e\sum_{j=1}^n \delta({\bm r}-{\bm r}_j),
\quad (e>0),
\end{align}
where the summation is taken over all $f$ electrons.
With this operator, it is natural to introduce the electric multipole operator as
\begin{align}
\hat{Q}_{pq}=-e\sum_j\int d{\bm r}\, \delta({\bm r}-{\bm r}_j) r^p_j Z_{pq}^*(\hat{\bm r}_j).
\label{e_mul_op}
\end{align}

A derivation of the magnetic multipole operator is more involved.
The detailed discussion is left in Appendix B, and we quote the result,
\begin{equation}
\hat{M}_{pq}=\mu_{\rm B}\sum_j
\int d{\bm r}\, \delta({\bm r}-{\bm r}_j) {\bm\nabla}\left( r_j^{p}Z_{pq}^{*}(\hat{\bm r}_j)\right)
\cdot
\left[\frac{2{\bm\ell}_j}{p+1}+2{\bm s}_j\right],
\label{m_mul_op}
\end{equation}
where $\mu_{\rm B}=-e\hbar/2mc$ is the Bohr magneton, and ${\bm\ell}_j$ and ${\bm s}_j$ are the orbital and the spin operators of $j$-th $f$ electron.

With these operators the classical multipole moments are given by the thermal average over $f$-electron states,
\begin{equation}
Q_{pq}=\Braket{\hat{Q}_{pq}}_f,
\quad\quad
M_{pq}=\Braket{\hat{M}_{pq}}_f.
\label{mul_av}
\end{equation}

\subsection{Spherical tensor and Stevens' operators}
In order to calculate systematically a matrix element of the multipole operators, let us consider the spherical (Racah) tensor operator\cite{Inui96}, $\hat{J}_{pq}$, which is defined by the $p$-th polynomial of the total angular momentum operator, $\hat{\bm J}=(\hat{J}_x,\hat{J}_y,\hat{J}_z)$.
The definition of the spherical tensor operator is
\begin{align}
&\hat{J}_{pp}=(-1)^p\sqrt{\frac{(2p-1)!!}{(2p)!!}}\left(\hat{J}_+\right)^p,
\notag\\
&\left[\hat{J}_-,\hat{J}_{pq}\right]=\sqrt{(p+q)(p-q+1)}\hat{J}_{pq-1},
\quad
(q<p),
\end{align}
where $\hat{J}_\pm=\hat{J}_x\pm i\hat{J}_y$.
We express the Wigner-Eckart theorem for the spherical tensor operator,
\begin{equation}
\Braket{JM'|\hat{J}_{pq}|JM}=(-1)^{J+M-p}
\begin{pmatrix}
J & J & p \\ -M' & M & q
\end{pmatrix}
\Braket{J||\hat{J}_p||J},
\end{equation}
where the parenthesis denotes the $3j$ symbol\cite{Landau81}, and the reduced matrix element of $\hat{J}_{pq}$ is given by
\begin{equation}
\Braket{J||\hat{J}_p||J}=\frac{1}{2^p}\sqrt{\frac{(2J+p+1)!}{(2J-p)!}}.
\label{jp_red}
\end{equation}
With use of the Wigner-Eckart theorem, we compute any matrix element of $\hat{J}_{pq}$ within a $J$ multiplet.

It is also possible to construct the hermite tensor operator\cite{Hutchings64} from $\hat{J}_{pq}$ with a similar linear combination as (\ref{real_rep}).
The explicit expression of the operator can be obtained by replacing $(x,y,z)$ in $r^pZ_{pq}^{\rm (c)}$, $r^pZ_{pq}^{\rm (s)}$ and $r^pZ_{p0}$ with the symmetrized product of $(\hat{J}_x,\hat{J}_y,\hat{J}_z)$, i.e.,
\begin{equation}
x^ky^mz^n\to
\frac{k!m!n!}{(k+m+n)!}\sum_{\cal P}{\cal P}\left(\hat{J}_x^k\hat{J}_y^m\hat{J}_z^n\right),
\end{equation}
where the summation is taken over possible permutations.
For instance, we obtain the operator form of $\hat{J}_{32}^{\rm (c)}$ by replacing
\begin{align}
r^3Z_{32}^{\rm (c)}&=\frac{\sqrt{15}}{2}z(x^2-y^2)
\notag\\ \quad\quad
&\to
\frac{\sqrt{15}}{6}\left[
\hat{J}_z(\hat{J}_x^2-\hat{J}_y^2)+(\hat{J}_x^2-\hat{J}_y^2)\hat{J}_z
+\hat{J}_x\hat{J}_z\hat{J}_x-\hat{J}_y\hat{J}_z\hat{J}_y
\right].
\end{align}
The point-group counterparts are obtained in a similar way.
We compute easily any matrix elements of $\hat{J}_{pq}$ and its variant with this prescription.

The so-called Stevens' operators\cite{Stevens52,Hutchings64} are often used in the CEF theory.
They are proportional to the hermite tensor operators as follows:
\begin{align}
&\hat{O}_{2}{}^{0} = 2\hat{J}_{20},\quad
\hat{O}_{2}{}^{2} = \frac{2}{\sqrt{3}}\hat{J}^{\rm (c)}_{22},
\notag\\
&\hat{O}_{4}{}^{0} = 8\hat{J}_{40},\quad
\hat{O}_{4}{}^{2} = \frac{4}{\sqrt{5}}\hat{J}^{\rm (c)}_{42},\quad
\hat{O}_{4}{}^{3} = \frac{4}{\sqrt{70}}\hat{J}_{43}^{\rm (c)},
\notag\\
&\hat{O}_{4}{}^{3}(s) = \frac{4}{\sqrt{70}}\hat{J}_{43}^{\rm (s)},\quad
\hat{O}_{4}{}^{4} = \frac{8}{\sqrt{35}}\hat{J}_{44}^{\rm (c)},\quad
\hat{O}_{4}{}^{4}(s) = \frac{8}{\sqrt{35}}\hat{J}_{44}^{\rm (s)},
\notag\\
&\hat{O}_{6}{}^{0} = 16\hat{J}_{60},\quad
\hat{O}_{6}{}^{2} = \frac{32}{\sqrt{210}}\hat{J}_{62}^{\rm (c)},\quad
\hat{O}_{6}{}^{3} = \frac{16}{\sqrt{210}}\hat{J}_{63}^{\rm (c)},
\notag\\&
\hat{O}_{6}{}^{4} = \frac{16}{3\sqrt{7}}\hat{J}_{64}^{\rm (c)},\quad
\hat{O}_{6}{}^{6} = \frac{32}{\sqrt{462}}\hat{J}_{66}^{\rm (c)}.
\end{align}
Note that the spherical and the point-group tensor operators are properly normalized, but the Stevens' operators are not.
The former is more appropriate for systematic calculation.

\subsection{Generalization of Stevens' operator equivalents}
Since $\hat{J}_{pq}$ is the $p$-th polynomial of the axial vector $\hat{J}_\alpha$ with time-reversal odd, the spherical tensor operator, $\hat{J}_{pq}$, has the even parity and the time reversal $(-1)^p$ Hereafter, the time-reversal symmetry of the tensor operator will be indicated by g(erade) and u(ngerade) in the subscript.
Moreover, the spherical tensor operator is transformed as $Z_{pq}(\hat{\bm r})$ by definition.
These symmetry properties are common with the even-rank electric and the odd-rank magnetic multipole operators.
Note that the even parity of the tensor operator differs from $(-1)^p$ parity of the spherical harmonics.
Therefore, according to the Wigner-Eckart theorem, we conclude that any matrix elements of the multipole operators are proportional to those of the corresponding spherical tensor operators.
We assume the presence of the inversion symmetry in what follows, then all of the relevant multipole operators are described by the spherical tensor operators.

\begin{figure}[tb]
\begin{center}
\includegraphics[width=8cm]{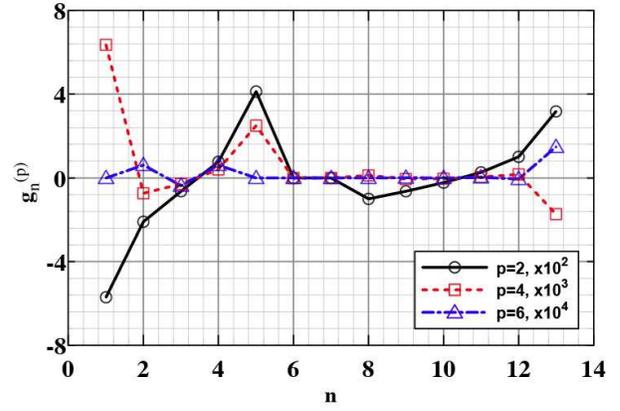}
\end{center}
\caption{(Color online) The even-rank Stevens' factors as a function of $n$.}
\label{g_even}
\end{figure}

\begin{figure}[tb]
\begin{center}
\includegraphics[width=8cm]{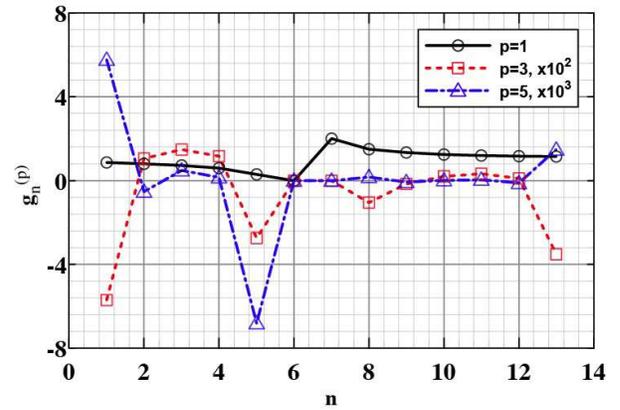}
\end{center}
\caption{(Color online) The odd-rank Stevens' factors as a function of $n$.}
\label{g_odd}
\end{figure}

Let us express any matrix element within a $J$ multiplet in an $f^n$ configuration as
\begin{subequations}
\begin{align}
&\Braket{nJM'|\hat{Q}_{pq}|nJM} = -e\Braket{r^p}g^{(p)}_n \Braket{JM'|\hat{J}_{pq}|JM},
\\
&\Braket{nJM'|\hat{M}_{pq}|nJM} = \mu_{\rm B}\Braket{r^{p-1}}g^{(p)}_n \Braket{JM'|\hat{J}_{pq}|JM},
\end{align}
\label{red_def}
\end{subequations}
where we have defined the radial average,
\begin{equation}
\Braket{r^k} = \int dr r^2 r^k R_f^2(r).
\end{equation}
The relativistic Hartree-Fock estimate for $\Braket{r^k}$ may be found in ref. [\citen{Freeman79}], for example.
Here, $g_n^{(p)}$ is the generalized Stevens' multiplicative factor, which is independent of $M$, $M'$ and $q$.
For the Hund's-rule ground multiplet, $^JL_{2S+1}$, in the Russell-Sanders ($LS$) scheme, $g_n^{(p)}$ is given in Table \ref{stevens}, and shown in Figs.\ref{g_even} and \ref{g_odd} as a function of the number of $f$ electrons.
The ratio of the orbital and the spin contributions to the magnetic multipole,
\begin{equation}
r_n^{(p)}=\frac{g_n^{(p)}(\text{orbital})}{g_n^{(p)}(\text{spin})},
\end{equation}
is shown in Fig. \ref{r_odd}.
The derivation of the generalized Stevens' factor is given in Appendix C.

Note that the even-rank $g_n^{(p)}$ is equivalent to the ordinary Stevens' factor\cite{Hutchings64}, and $g_n^{(1)}$ is nothing but the Land\'e's $g$ factor\cite{Inui96,Hutchings64}, i.e.,
\begin{equation}
g_n^{(1)}=g_J,
\end{equation}
and
\begin{equation}
g_n^{(2)}=\theta_2=\alpha_J,
\quad
g_n^{(4)}=\theta_4=\beta_J,
\quad
g_n^{(6)}=\theta_6=\gamma_J.
\end{equation}

\begin{figure}[tb]
\begin{center}
\includegraphics[width=8cm]{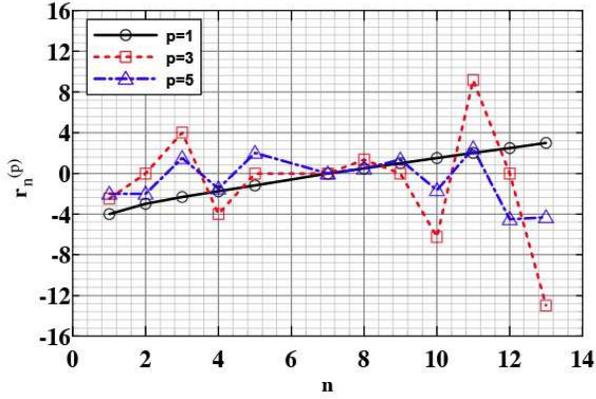}
\end{center}
\caption{(Color online) The ratio of the orbital and the spin contributions to the magnetic multipoles.}
\label{r_odd}
\end{figure}

Since the scalar and the vector potentials are one-body fields, the multipole operators interacting with them are represented by one-body operators as shown in (\ref{e_mul_op}) and (\ref{m_mul_op}).
Due to the selection rule (\ref{selection_sh}) of the spherical harmonics, the Stevens' factors more than rank $p>2\ell=6$ must vanish.
In view of a quantum-mechanical transition, multipole operators with $p>6$, e.g., $\hat{\cal O}_{8}{}^{8}\sim (\Ket{4,+4}\Bra{4,-4}+{\rm h.c.})$, could exist for $J>3$.
However, such operators with $p>6$ are represented by more-than two-body operators, and do not couple with one-body potentials.

Let us express an arbitrary state within a $J$ multiplet,
\begin{equation}
\Ket{\gamma}=\sum_MU_{M\gamma}\Ket{JM}.
\end{equation}
Note that the unitary matrix $U_{M\gamma}$ could be complex.
With the operator-equivalent method, the classical multipole moments are given by
\begin{equation}
\frac{Q_{pq}}{-e\Braket{r^p}},\,\frac{M_{pq}}{\mu_{\rm B}\Braket{r^{p-1}}}=g_n^{(p)}\Braket{J||\hat{J}_p||J}\sum_\gamma {\cal W}_{pq}(\gamma)\frac{e^{-\beta E_\gamma}}{Z_f},
\label{mul_val}
\end{equation}
where we have defined the weight function for a $\gamma$ state as
\begin{equation}
{\cal W}_{pq}(\gamma)=\sum_{MM'}(-1)^{J+M-p}
\begin{pmatrix}
J & J & p \\ -M' & M & q
\end{pmatrix}
U_{M'\gamma}^*U_{M\gamma}.
\end{equation}
Note that in the case of $U_{M\gamma}=\delta_{M\gamma}$, ${\cal W}_{pq}(\gamma)=0$ unless $q=0$.

\section{Visualization of Wave Function}
In this section, we consider visualization for the charge density $\rho({\bm r};\gamma)$ and the magnetic charge density $\rho_{\rm m}({\bm r};\gamma)$ of $f$ electrons in a particular state $\Ket{\gamma}$.
From (\ref{e_multipole1}) and (\ref{mul_av}), we may obtain the relation
\begin{equation}
\Braket{\gamma | \hat{Q}_{pq} | \gamma}=\int d{\bm r}\, r^pZ_{pq}^*(\hat{\bm r})\rho({\bm r};\gamma).
\label{e_multipole_part}
\end{equation}
Suppose that we write $\rho({\bm r};\gamma)$ in the separable form,
\begin{equation}
\rho({\bm r};\gamma)=-eR_f^2(r)\rho_{\rm e}(\hat{\bm r};\gamma)/4\pi.
\end{equation}
Substituting this into (\ref{e_multipole_part}) and using the completeness of the spherical harmonics,
\begin{equation}
\sum_{pq}\frac{2p+1}{4\pi}Z_{pq}(\hat{\bm r}')Z_{pq}^*(\hat{\bm r})=\delta(\hat{\bm r}-\hat{\bm r}'),
\end{equation}
we obtain the angle dependence of the charge density,
\begin{equation}
\rho_{\rm e}(\hat{\bm r};\gamma)=\sum_{p=0}^\infty\sum_{q=-p}^p(2p+1)\frac{
\Braket{\gamma | \hat{Q}_{pq} | \gamma}}{-e\Braket{r^p}}Z_{pq}(\hat{\bm r}). 
\end{equation}
This expression is equivalent to that obtained by Walter\cite{Walter86}.

Similarly, supposing that
\begin{equation}
\rho_{\rm m}({\bm r};\gamma)=\mu_{\rm B}\frac{R_f^2(r)}{r}\rho_{\rm m}(\hat{\bm r};\gamma)/4\pi,
\end{equation}
and (\ref{mm1}), we obtain the angle dependence of the magnetic charge density,
\begin{equation}
\rho_{\rm m}(\hat{\bm r};\gamma)=\sum_{p=0}^\infty\sum_{q=-p}^p(2p+1)\frac{
\Braket{\gamma | \hat{M}_{pq} | \gamma}}{\mu_{\rm B}\Braket{r^{p-1}}}Z_{pq}(\hat{\bm r}). 
\end{equation}

Using the operator equivalents (\ref{red_def}), we express the angle dependences in the common form,
\begin{align}
&\rho_{\rm e,m}(\hat{\bm r};\gamma)=
\sum_{p=0}^6(2p+1)g_n^{(p)}\Braket{J || \hat{J}_{p} || J}\sum_{q=-p}^p{\cal W}_{pq}(\gamma)Z_{pq}(\hat{\bm r}),
\label{rhoem}
\end{align}
where the summation for $p$ is taken over even (odd) integers for $\rho_{\rm e}$ ($\rho_{\rm m}$).
This expression is useful to visualize a wave function for a $\gamma$ state.
The presentation of the charge density is similar to that in refs. [\citen{Walter86,Sievers82}].
Namely, the radius $R_\gamma(\hat{\bm r})$ to the surface of the 3-dimensional plot is defined as
\begin{equation}
R_\gamma(\hat{\bm r})=\left[\rho_{\rm e}(\hat{\bm r};\gamma)\right]^\alpha,
\end{equation}
in which $\alpha=1$ is chosen to emphasize the gradation of the charge density, although $\alpha=1/3$ is natural to yield that the encircled volume becomes the total charge.
The surface-color map is used to represent the magnetic charge density $\rho_{\rm m}(\hat{\bm r};\gamma)$, which is normalized to hold the range of the distribution in $[-1,1]$.

\section{Treatments for Multipole Exchange Systems}
\subsection{The exchange model in a crystal}
As was mentioned in \S2.4, the point-group representation is appropriate in the case of large CEF splitting.
When one of CEF multiplets or a bunch of CEF multiplets with small splittings dominate low-energy physics, we consider a multipole exchange system within the relevant CEF states.
In the restricted basis with the (pseudo) degeneracy $d$, the multipole operators in the point-group representation become reducible, and some of them are proportional with each other.

Meanwhile, in the view of the quantum-mechanical variables, we require $d^2$ independent operators to expand the restricted manifold (one of these is the identity operator).
Since $d^2=\sum_{p=0}^{d-1}(2p+1)$, we formally assign the spherical tensor operators up to rank $d-1$ to the $d^2$ independent operators\cite{Shiina97}.
Thus, such mathematically independent operator is called the multipole operator as well.

Keeping this consideration in mind, we denote the $d^2-1$ independent operators at the site $i$ as $\hat{X}_i^\alpha$ ($\alpha=1,2,\cdots d^2-1$), except the identity operator denoted as $\hat{1}_i$.
The operator $\hat{X}_i^\alpha$ is hermite, traceless and is normalized as
\begin{equation}
\frac{1}{d}{\rm Tr}_i\left(\hat{X}_i^\alpha\hat{X}_i^\beta\right)=\delta_{\alpha\beta}.
\label{orthonormality}
\end{equation}
In terms of $\hat{X}_i^\alpha$, we write down a generalized exchange model with uniform external fields,
\begin{equation}
{\cal H}_f=-\frac{1}{2}\sum_{ij}\sum_{\alpha\beta}D_{ij}^{\alpha\beta}\hat{X}_i^\alpha\hat{X}_j^\beta-\sum_{i\alpha}\hat{X}_i^\alpha h^\alpha,
\label{exchange_model}
\end{equation}
where we assume complete degeneracy of CEF states for simplicity, but a generalization to pseudo-degeneracy is straightforward.
Note that possible differences in normalization of the multipole operators are absorbed in the definition of the exchange coupling $D_{ij}^{\alpha\beta}$.
However, it should be emphasized that the relation between $\hat{X}_i^\alpha$ and the multipole operators, $\hat{Q}_{pq}$ and $\hat{M}_{pq}$, is vital in evaluating an effect of the multipoles through electromagnetic probes.
The non-vanishing combination of $\alpha$ and $\beta$ in $D_{ij}^{\alpha\beta}$ may be obtained by symmetry consideration with respect to an interacting bond $i$-$j$\cite{Sakai03}.
The second term represents coupling with uniform external fields such as magnetic field, uniaxial strain, and so on.

As was mentioned in \S3.3, the multipole operators with $p>6$ consist of more than two-body operators.
The exchange couplings are expected to be small for such operators because the origin of $D_{ij}^{\alpha\beta}$ involves higher-order exchange processes to transfer more than two electron states simultaneously.
To the contrary, $D_{ij}^{\alpha\beta}$ for $p\le 6$ could be the same order in magnitude.
This is because the origin of the coupling is the RKKY and/or the superexchange mechanism in which the matrix elements and the intermediate energies have similar strength.
Especially, when a virtual process through a featureless state such as $f^0$ configuration dominates, all of the coupling strength turns to be coincident with each other\cite{Shiba99}.

Since the operators $\hat{X}_i^\alpha$ span the restricted manifold, the product of the operators can be expanded as
\begin{equation}
\hat{X}_i^\alpha\hat{X}_i^\beta=\sum_\gamma \left(if_{\alpha\beta\gamma}+g_{\alpha\beta\gamma}\right)\hat{X}_i^\gamma+\delta_{\alpha\beta}\hat{1}_i,
\label{xx_expansion}
\end{equation}
where the symmetric and anti-symmetric structure constants are calculated from the definition of $\hat{X}_i^{\alpha}$ as
\begin{subequations}
\begin{align}
&g_{\alpha\beta\gamma}=\frac{1}{2d}{\rm Tr}_i\,\left(\left[\hat{X}_i^\alpha\hat{X}_i^\beta+\hat{X}_i^\beta\hat{X}_i^\alpha\right]\hat{X}_i^\gamma\right),
\\
&if_{\alpha\beta\gamma}=\frac{1}{2d}{\rm Tr}_i\,\left(\left[\hat{X}_i^\alpha\hat{X}_i^\beta-\hat{X}_i^\beta\hat{X}_i^\alpha\right]\hat{X}_i^\gamma\right).
\end{align}
\label{structure_const}
\end{subequations}
The symmetry property of the system is completely determined by the structure constants.
In the case of the Pauli matrices, $\hat{X}_i^\alpha=\hat{\sigma}_i^\alpha$, we have $g_{\alpha\beta\gamma}=0$ and $f_{\alpha\beta\gamma}=\epsilon_{\alpha\beta\gamma}$, where $\epsilon_{\alpha\beta\gamma}$ is the anti-symmetric (Levi-Civita) symbol.

\subsection{The RPA susceptibility}
Let us consider the static susceptibility of the multipoles within RPA.
The second-order phase transition from a disorder phase is then determined by the divergence of the susceptibility in the mean-field approximation.
For this purpose, we add a coupling with fictitious fields to the exchange Hamiltonian,
\begin{equation}
{\cal H}={\cal H}_f-\sum_{i\alpha}\hat{X}_i^\alpha\phi_i^\alpha.
\end{equation}
We divide the thermal average of the multipole operator into two parts,
\begin{equation}
\Braket{X_i^\alpha}=\Braket{\hat{X}_i^\alpha}_f+\Braket{\hat{X}_i^\alpha}_\phi,
\end{equation}
where the second term is the induced moment being proportional to the infinitesimally small field $\phi_i^\alpha$.
Note that $\Braket{\hat{X}_i^\alpha}_f$ could be finite in the presence of external fields, otherwise they must vanish in the disorder phase.

The mean-field Hamiltonian is
\begin{equation}
{\cal H}_{\rm MF}=\sum_i\left[ {\cal H}_f(i)-\sum_\alpha \hat{X}_i^\alpha\lambda_i^\alpha \right],
\end{equation}
where we define the mean-field Hamiltonian without $\phi_i^\alpha$ fields,
\begin{equation}
{\cal H}_f(i)=-\sum_\alpha\left(h^\alpha+\sum_{j\beta}D_{ij}^{\alpha\beta}\Braket{\hat{X}_j^\beta}_f\right)\hat{X}_i^\alpha,
\end{equation}
and the effective infinitesimal field,
\begin{equation}
\lambda_i^\alpha=\phi_i^\alpha+\sum_{j\beta}D_{ij}^{\alpha\beta}\Braket{\hat{X}_j^\beta}_\phi.
\end{equation}
The self-consistent equation for $h^\alpha\ne0$ and $\phi_i^\alpha=0$ is given by
\begin{equation}
\Braket{\hat{X}_i^\alpha}_f=\sum_m f_m\Braket{m| \hat{X}_i^\alpha |m},
\quad
f_m\equiv\frac{e^{-\beta E_m}}{\sum_me^{-\beta E_m}},
\end{equation}
where ${\cal H}_f(i)\Ket{m}=E_m\Ket{m}$ and $\Braket{\hat{X}_i^\alpha}_f$ is independent of the site in the disorder phase.

According to the linear-response theory, we have the local susceptibility for $\phi_i^\alpha=0$,
\begin{align}
&\chi_{\rm loc}^{\alpha\beta}=\sum_{mn}f_m\frac{1-e^{-\beta(E_n-E_m)}}{E_n-E_m}
\Braket{m|\hat{X}_i^\alpha|n}
\Braket{n|\hat{X}_i^\beta|m}
\notag\\&\quad\quad\quad\quad\quad\quad\quad\quad\quad\quad
-\beta\Braket{\hat{X}_i^\alpha}_f
\Braket{\hat{X}_i^\beta}_f.
\end{align}
Note that in the absence of the external fields, $h^\alpha=0$, we have the Curie law, $\chi_{\rm loc}^{\alpha\beta}(h^\alpha=0)=\beta\delta_{\alpha\beta}$.
Using $\chi_{\rm loc}^{\alpha\beta}$, we obtain
\begin{equation}
\Braket{\hat{X}_i^\alpha}_\phi=\sum_\beta\chi_{\rm loc}^{\alpha\beta}\lambda_i^\beta=\sum_\gamma\chi_{\rm loc}^{\alpha\gamma}\left[\phi_i^\gamma+\sum_{k\delta}D_{ik}^{\gamma\delta}\Braket{\hat{X}_k^\delta}_\phi\right].
\end{equation}
By the definition of the susceptibility, $\chi_{ij}^{\alpha\beta}=\partial\Braket{\hat{X}_i^\alpha}_\phi/\partial \phi_j^\beta|_{\phi=0}$,
we have the relation,
\begin{equation}
\chi_{ij}^{\alpha\beta}=\sum_\gamma\chi_{\rm loc}^{\alpha\gamma}\left[\delta_{ij}\delta_{\gamma\beta}+\sum_{k\delta}D_{ik}^{\gamma\delta}\chi_{kj}^{\delta\beta}\right].
\end{equation}
With the Fourier transformation, we finally obtain the RPA susceptibility,
\begin{equation}
\chi_{\rm RPA}^{\alpha\beta}({\bm q})=\sum_\gamma\left[\hat{1}-\hat{\chi}_{\rm loc}\hat{D}({\bm q})\right]^{-1}_{\alpha\gamma}\chi_{\rm loc}^{\gamma\beta},
\label{rpa_sus}
\end{equation}
where $D^{\alpha\beta}({\bm q})=\sum_n e^{-i{\bm q}\cdot{\bm r}_n}D_{n0}^{\alpha\beta}$.
In the case of diagonal coupling $D^{\alpha\beta}=D^\alpha\delta_{\alpha\beta}$ and $h^\alpha=0$, we have the simple Curie-Weiss susceptibility,
\begin{equation}
\chi_{\rm RPA}^{\alpha\beta}({\bm q})=\frac{1}{T-D^\alpha({\bm q})}\delta_{\alpha\beta}.
\end{equation}

The second-order phase transition is determined by
\begin{equation}
{\rm det}\,\left[ \delta_{\alpha\beta}-\sum_\gamma\chi_{\rm loc}^{\alpha\gamma}D^{\gamma\beta}({\bm q})\right]=0,
\label{ev_eq}
\end{equation}
at ${\bm q}={\bm Q}$ with the maximum $T_c$.
The ratio of the order parameters just below $T_c$ is determined by the eigenvector of the matrix in the eigenvalue equation.

\subsection{The Ginzburg-Landau free energy}
The GL free-energy expansion is useful to elucidate systematically an entanglement of the multipoles.
Here we derive the general expression of the GL free energy of the exchange model for $h^\alpha=0$.
Let us consider the one-body trial Hamiltonian,
\begin{equation}
{\cal H}_0=-\sum_{i\alpha}\hat{X}_i^\alpha\psi_i^\alpha.
\end{equation}
Then, the exact free energy is upper-bounded by the Feynman inequality\cite{Feynman98},
\begin{equation}
{\cal F}\le {\cal F}_{\rm tr}\equiv {\cal F}_0+\Braket{ {\cal H}_f-{\cal H}_0 }_0,
\end{equation}
where $\Braket{\cdots}_0$ is the thermal average with respect to ${\cal H}_0$ and ${\cal F}_0=-\beta^{-1}\ln\,{\rm Tr}e^{-\beta{\cal H}_0}$.
Minimizing ${\cal F}_{\rm tr}$ with respect to $\psi_i^\alpha$, we obtain the best Hamiltonian within the one-body approximation.
This type of variational treatment is equivalent to the mean-field theory.

The thermal average is easily evaluated as
\begin{align}
&\Braket{ {\cal H}_f-{\cal H}_0 }_0=-\frac{1}{2}\sum_{ij}\sum_{\alpha\beta}D_{ij}^{\alpha\beta}
\Braket{ \hat{X}_i^\alpha}_0\Braket{ \hat{X}_j^\beta}_0
+\sum_{i\alpha}\Braket{ \hat{X}_i^\alpha}_0\psi_i^\alpha.
\end{align}
By using $\partial\Braket{ \hat{X}_i^\alpha}_0/\partial\psi_k^\gamma=\beta\delta_{ik}\delta_{\alpha\gamma}$,
 the stationary condition yields
\begin{equation}
\overline{\psi}_i^\alpha=\sum_{j\beta}D_{ij}^{\alpha\beta}\Braket{\hat{X}_j^\beta}_0\biggl|_{\psi_i^\alpha=\overline{\psi}_i^\alpha},
\label{mfp}
\end{equation}
which plays a role of the self-consistent equation.
Eliminating the order parameter $X_i^\alpha\equiv\Braket{\hat{X}_i^\alpha}_0\biggl|_{\psi_i^\alpha=\overline{\psi}_i^\alpha}$ with the stationary condition, we obtain the best trial free energy as
\begin{align}
\overline{\cal F}_{\rm tr}=\frac{1}{2}\sum_{ij}\sum_{\alpha\beta}\left(D^{-1}\right)_{ij}^{\alpha\beta}
\overline{\psi}_i^\alpha\overline{\psi}_j^\beta
-\frac{1}{\beta}\sum_i\ln\left[
{\rm Tr}_ie^{\beta\sum_\alpha \hat{X}_i^\alpha \overline{\psi}_i^\alpha}\right].
\end{align}

Now, we express it in terms of the order parameter $X_i^\alpha$.
Expanding $X_i^\alpha$ up to $\overline{\psi}^3$ with use of (\ref{xx_expansion}), we have
\begin{align}
&X_i^\alpha\sim\beta\overline{\psi}_i^\alpha+\frac{\beta^2}{2}\sum_{\beta\gamma}g_{\alpha\beta\gamma}\overline{\psi}_i^\beta\overline{\psi}_i^\gamma
\notag\\&\quad\quad\quad\quad
+\frac{\beta^3}{6}\sum_{\beta\gamma\delta}\left(L_{\alpha\beta\gamma\delta}-3\delta_{\alpha\beta}\delta_{\gamma\delta}\right)
\overline{\psi}_i^\beta\overline{\psi}_i^\gamma\overline{\psi}_i^\delta,
\label{x_exp}
\end{align}
where we have introduced
\begin{equation}
L_{\alpha\beta\gamma\delta}=\sum_\xi g_{\alpha\beta\xi}g_{\gamma\delta\xi}+\delta_{\alpha\beta}\delta_{\gamma\delta}.
\end{equation}
The converse relation of (\ref{x_exp}) is obtained by a recursive expansion as
\begin{equation}
\beta\overline{\psi}_i^\alpha\sim X_i^\alpha-\frac{1}{2}\sum_{\beta\gamma}g_{\alpha\beta\gamma}X_i^\beta X_i^\gamma+\frac{1}{3}\sum_{\beta\gamma\delta}
L_{\alpha\beta\gamma\delta}
X_i^\beta X_i^\gamma X_i^\delta.
\end{equation}
Using this relation and (\ref{mfp}), we finally obtain the GL free energy up to 4th order (except the $-TN\ln d$ term),
\begin{align}
&{\cal F}_{\rm GL}\equiv\overline{\cal F}_{\rm tr}=
\overline{\cal F}_{\rm tr}+\sum_{i\alpha}\left(
\overline{\psi}_i^\alpha-\sum_{j\beta}D_{ij}^{\alpha\beta}X_j^\beta
\right)
X_i^\alpha
\notag\\&\quad\quad
=\frac{1}{2}\sum_{ij\alpha\beta}\left(T\delta_{ij}\delta_{\alpha\beta}- D_{ij}^{\alpha\beta}\right)X_i^\alpha X_j^\beta-\frac{T}{6}\sum_{i\alpha\beta\gamma} g_{\alpha\beta\gamma}X_i^\alpha X_i^\beta X_i^\gamma
\notag\\&\quad\quad\quad\quad\quad
+\frac{T}{12}\sum_{i\alpha\beta\gamma\delta} L_{\alpha\beta\gamma\delta} X_i^\alpha X_i^\beta X_i^\gamma X_i^\delta
+{\cal O}(X^5).
\label{gl_fe}
\end{align}
In the case of $\hat{X}_i^\alpha=\hat{\sigma}_i^\alpha$ and $D_{ij}^{\alpha\beta}=J_{ij}\delta_{\alpha\beta}$, we recover the GL free energy of the SU(2) Heisenberg model (we denote $m_i^\alpha=X_i^\alpha$),
\begin{equation}
{\cal F}_{\rm GL}=
\frac{1}{2}\sum_{ij}\left(T\delta_{ij}-J_{ij}\right){\bm m}_i\cdot {\bm m}_j
+\frac{T}{12}\sum_i({\bm m}_i\cdot{\bm m}_i)^2.
\end{equation}

The self-consistent equation (\ref{mfp}) reduces to the condition $\partial {\cal F}_{\rm GL}/\partial X_i^\alpha=0$, and it is given in the GL expansion as
\begin{align}
&\sum_{j\beta}(T\delta_{ij}\delta_{\alpha\beta}-D_{ij}^{\alpha\beta})X_j^\beta-\frac{T}{2}\sum_{\beta\gamma}g_{\alpha\beta\gamma}X_i^\beta X_j^\gamma
\notag\\&\quad\quad\quad\quad
+\frac{T}{3}\sum_{\beta\gamma\delta}L_{\alpha\beta\gamma}X_i^\beta X_i^\gamma X_i^\delta=0.
\end{align}
The fluctuation from the stationary is related to the susceptibility.
Namely, we replace $X_i^\alpha\to X_i^\alpha+\delta X_i^\alpha$ and retain terms up to 2nd order in $\delta X_i^\alpha$, then we obtain the deviation of the free energy,
\begin{align}
&\delta \overline{\cal F}_{\rm GL}=\frac{1}{2}\sum_{ij\alpha\beta}\left(\chi^{-1}\right)_{ij}^{\alpha\beta}\delta X_i^\alpha \delta X_j^\beta,
\\&
\left(\chi^{-1}\right)_{ij}^{\alpha\beta}=
T\delta_{ij}\delta_{\alpha\beta}-D_{ij}^{\alpha\beta}
\notag\\&\quad\quad
-T\left[
\sum_\gamma g_{\alpha\beta\gamma} X_i^\gamma
-\frac{1}{3}\sum_{\gamma\delta}\left(L_{\alpha\beta\gamma\delta}+2L_{\alpha\gamma\beta\delta}\right)X_i^\gamma X_i^\delta\right]
\delta_{ij}.
\label{sus_order}
\end{align}
Since $X_i^\alpha=0$ in the disorder phase, we recover the RPA susceptibility (\ref{rpa_sus}) for $h^\alpha=0$.
On the other hand, the mode mixing arises through the 3rd and 4th-order couplings in the ordered phase, $X_i^\alpha\ne0$.

The simplest non-trivial entanglement of the multipoles arise from the 3rd-order coupling, $g_{\alpha\beta\gamma}$.
The coupling conserves the momenta of the order parameters.
Namely, the entanglement comes from uniqueness of the local symmetry-breaking wave function.
When a spontaneous order occurs for $X_i^\alpha$, the mode mixing takes place between $X_i^\beta$ and $X_i^\gamma$.
If two of three multipoles are equivalent, e.g. $\alpha=\beta$, $X_i^\gamma$ is induced eventually.
Since the free energy is even under time reversal, the 3rd-order term should consist of three electric multipoles or two magnetic and one electric multipoles.
Consequently, the magnetic multipole is the primary order parameter while the electric multipole is secondary when both multipoles coexist in the ordered phase.

\section{The example with Ce$_{x}$La$_{1-x}$B$_6$}
In this section, we illustrate a use of the previous arguments with Ce$_{1-x}$La$_x$B$_6$ as an example.
The trivalent Ce ion is placed in the cubic crystal field $O_h$, and $J=5/2$ multiplet splits into $\Gamma_7$ doublet and $\Gamma_8$ quartet.
The wave functions are given by
\begin{subequations}
\begin{align}
&\Ket{\Gamma_7;\pm}=\sqrt{\frac{1}{6}}\Ket{\pm\frac{5}{2}}-\sqrt{\frac{5}{6}}\Ket{\mp\frac{3}{2}},
\\
&
\Ket{\Gamma_8;a\pm}=\sqrt{\frac{5}{6}}\Ket{\pm\frac{5}{2}}+\sqrt{\frac{1}{6}}\Ket{\mp\frac{3}{2}},
\\
&
\Ket{\Gamma_8;b\pm}=\Ket{\pm\frac{1}{2}}.
\end{align}
\end{subequations}
$\Gamma_7$ doublet consists of the time-reversal pair, while $\Gamma_8$ quartet has an additional orbital degrees of freedom.
The splitting between $\Gamma_8$ ground state and $\Gamma_7$ excited state is about 500K, so that it provides an ideal quartet system in low temperatures.
There are four phases at most in $H$-$T$ phase diagram, which are called as I-IV\cite{Tayama97}.

Within $\Gamma_7$ doublet, we decompose the direct product of the basis as $\Gamma_7\otimes\Gamma_7=\Gamma_1\oplus\Gamma_4$, which indicates that two types of the multipole operators including the identity one are active in $\Gamma_7$ doublet.
We express a set of independent operators in terms of the $2\times2$ Pauli matrices $\hat{\sigma}_\gamma$ and the identity matrix $\hat{1}$.
They are related to the spherical tensor operators as
\begin{subequations}
\begin{align}
&\Gamma_{1g}:\,\,
\hat{1}=\hat{J}_{0,1g,1}=-\frac{1}{5\sqrt{21}}\hat{J}_{4,1g,1},
\\
&\Gamma_{4u}:\,\,
\hat{X}^{1,2,3}=\hat{\sigma}_\gamma=-\frac{6}{5}\hat{J}_{1,4u,\gamma}=\frac{1}{10}\hat{J}_{3,4u,\gamma}
\notag\\&\quad\quad\quad\quad\quad\quad\quad
=\frac{4}{65}\hat{J}_{5,4au,\gamma}
=-\frac{4}{15\sqrt{35}}\hat{J}_{5,4bu,\gamma},
\end{align}
\end{subequations}
where the cubic tensor operator is expressed as $\hat{J}_{p,\Gamma(g,u),\gamma}$.
Note that the operators belonging to the same $\Gamma$ with different ranks are reducible in the subspace.

On the other hand, the decomposition for $\Gamma_8$ quartet\cite{Shiina97} is $\Gamma_8\otimes\Gamma_8=\Gamma_1\oplus\Gamma_2\oplus\Gamma_3\oplus2\Gamma_4\oplus2\Gamma_5$.
A set of independent operators in $\Gamma_8$ quartet is given by
\begin{subequations}
\begin{align}
&\Gamma_{1g}:\,\,
\hat{1}=\hat{J}_{0,1g,1}=\frac{2}{5\sqrt{21}}\hat{J}_{4,1g,1},
\\
&\Gamma_{2u}:\,\,
\hat{X}^{1}=\hat{\tau}_y=\frac{2}{9\sqrt{5}}\hat{J}_{3,2u,1},
\\
&\Gamma_{3g}:\,\,
\hat{X}^{2,3}=(\hat{\tau}_z,\hat{\tau}_x)=\frac{1}{4}\hat{J}_{2,3g,\gamma}=-\frac{2}{5\sqrt{15}}\hat{J}_{4,3g,\gamma},
\\
&\Gamma_{4Au}:\,\,
\hat{X}^{4,5,6}=\hat{\sigma}_\gamma=\frac{14}{15}\hat{J}_{1,4u,\gamma}-\frac{4}{45}\hat{J}_{3,4u,\gamma}
\notag\\&\quad\quad\quad\quad\quad\quad\quad\quad
=\frac{2}{75}\hat{J}_{5,4au,\gamma}+\frac{2}{9\sqrt{35}}\hat{J}_{5,4bu,\gamma},
\\
&\Gamma_{4Bu}:\,\,
\hat{X}^{7,8,9}=(\hat{\eta}_+\hat{\sigma}_x,\hat{\eta}_-\hat{\sigma}_y,\hat{\tau}_z\hat{\sigma}_z)=-\frac{2}{15}\hat{J}_{1,4u,\gamma}+\frac{7}{45}\hat{J}_{3,4u,\gamma}
\notag\\&\quad\quad\quad\quad\quad
=-\frac{2}{75}\hat{J}_{5,4au,\gamma}+\frac{14}{45\sqrt{35}}\hat{J}_{5,4bu,\gamma},
\\
&\Gamma_{5u}:\,\,
\hat{X}^{10,11,12}=(\hat{\zeta}_+\hat{\sigma}_x,\hat{\zeta}_-\hat{\sigma}_y,\hat{\tau}_x\hat{\sigma}_z)
\notag\\&\quad\quad\quad\quad\quad\quad
=\frac{1}{3\sqrt{5}}\hat{J}_{3,5u,\gamma}
=\frac{4}{15\sqrt{35}}\hat{J}_{5,5u,\gamma},
\label{mmul5u}
\\
&\Gamma_{5g}:\,\,
\hat{X}^{13,14,15}=\hat{\tau}_y\hat{\sigma}_\gamma=\hat{J}_{2,5g,\gamma}=\frac{1}{5\sqrt{15}}\hat{J}_{4,5g,\gamma},
\end{align}
\end{subequations}
where $\hat{\sigma}_\gamma$ and $\hat{\tau}_\gamma$ are $4\times4$ matrices acting on the time-reversal and the orbital pairs respectively, and we have defined
\begin{subequations}
\begin{align}
&\hat{\eta}_\pm=-\frac{1}{2}\left(\hat{\tau}_x\pm\sqrt{3}\hat{\tau}_z\right),\\
&\hat{\zeta}_\pm=-\frac{1}{2}\left(\hat{\tau}_z\mp\sqrt{3}\hat{\tau}_x\right).
\end{align}
\end{subequations}
Note that the magnetic dipole moment belongs to $\Gamma_{4u}$ irreducible representation, and is expressed as
\begin{equation}
\hat{M}_{\gamma}=g_1^{(1)}\hat{J}_{1,4u,\gamma}=\hat{X}^{4,5,6}+\frac{4}{7}\hat{X}^{7,8,9}.
\end{equation}
The operators $\hat{X}^\alpha$ satisfy the orthonormality\cite{xortho}, (\ref{orthonormality}).
The symmetric and the anti-symmetric structure constants are summarized in Tables \ref{symconst} and \ref{asymconst}.

\begin{table}[tb]
\caption{The symmetric structure constants, $g_{\alpha\beta\gamma}$.}
\label{symconst}
\begin{tabular}{lclclc} \hline
$\alpha,\beta,\gamma$ & $g_{\alpha\beta\gamma}$ & $\alpha,\beta,\gamma$ & $g_{\alpha\beta\gamma}$ & $\alpha,\beta,\gamma$ & $g_{\alpha\beta\gamma}$ \\ \hline
1,4,13   & $1$           & 1,5,14   & $1$           & 1,6,15   & $1$ \\
2,4,7    & $-1/2$        & 2,4,10   & $\sqrt{3}/2$  & 2,5,8    & $1/2$ \\
2,5,11   & $-\sqrt{3}/2$ & 2,6,9    & $1$           & 3,4,7    & $-\sqrt{3}/2$ \\
3,4,10   & $-1/2$        & 3,5,8    & $-\sqrt{3}/2$ & 3,5,11   & $-1/2$ \\
3,6,12   & $1$           & 7,8,15   & $-\sqrt{3}/2$ & 7,9,14   & $\sqrt{3}/2$ \\
7,11,15  & $1/2$         & 7,12,14  & $-1/2$        & 8,9,13   & $-\sqrt{3}/2$ \\
8,10,15  & $1/2$         & 8,12,13  & $-1/2$        & 9,10,14  & $1/2$ \\
9,11,13  & $-1/2$        & 10,11,15 & $\sqrt{3}/2$  & 10,12,14 & $\sqrt{3}/2$ \\
11,12,13 & $\sqrt{3}/2$ \\ \hline
\end{tabular}
\end{table}

\begin{table}[tb]
\caption{The anti-symmetric structure constants, $f_{\alpha\beta\gamma}$.}
\label{asymconst}
\begin{tabular}{lclclc} \hline
$\alpha,\beta,\gamma$ & $f_{\alpha\beta\gamma}$ & $\alpha,\beta,\gamma$ & $f_{\alpha\beta\gamma}$ & $\alpha,\beta,\gamma$ & $f_{\alpha\beta\gamma}$ \\ \hline
1,2,3    & $1$           & 1,7,10   & $1$           & 1,8,11   & $-1$ \\
1,9,12   & $1$           & 2,7,13   & $-\sqrt{3}/2$ & 2,8,14   & $-\sqrt{3}/2$ \\
2,10,13  & $-1/2$        & 2,11,14  & $-1/2$        & 2,12,15  & $1$ \\
3,7,13   & $1/2$         & 3,8,14   & $-1/2$        & 3,9,15   & $-1$ \\
3,10,13  & $-\sqrt{3}/2$ & 3,11,14  & $\sqrt{3}/2$  & 4,5,6    & $1$ \\
4,8,9    & $1/2$         & 4,8,12   & $-\sqrt{3}/2$ & 4,9,11   & $\sqrt{3}/2$ \\
4,11,12  & $-1/2$        & 4,14,15  & $1$           & 5,7,9    & $1/2$ \\
5,7,12   & $\sqrt{3}/2$  & 5,9,10   & $\sqrt{3}/2$  & 5,10,12  & $1/2$ \\
5,13,15  & $-1$          & 6,7,8    & $1/2$         & 6,7,11   & $\sqrt{3}/2$ \\
6,8,10   & $-\sqrt{3}/2$ & 6,10,11  & $-1/2$        & 6,13,14  & $1$ \\ \hline
\end{tabular}
\end{table}

\begin{figure}[tb]
\begin{center}
\includegraphics[width=6cm]{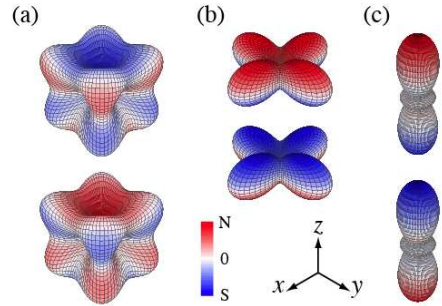}
\end{center}
\caption{(Color online) The charge and the magnetic charge densities for Ce$^{3+}$ viewed from [111], (a) $\Gamma_7\pm$, (b) $\Gamma_{8a}\pm$ and (c) $\Gamma_{8b}\pm$.}
\label{wf_ce}
\end{figure}

From (\ref{rhoem}), the angle dependences of $\Gamma_7$ wave functions are expressed in terms of the cubic harmonics, $Z_{p,\Gamma,\gamma}(\hat{\bm r})$,
\begin{subequations}
\begin{align}
&\rho_{\rm e}(\hat{\bm r};\Gamma_7\pm)=g_1^{(0)}-45\sqrt{21}g_1^{(4)}Z_{4,1,1}(\hat{\bm r}),
\\
&\rho_{\rm m}(\hat{\bm r};\Gamma_7\pm)=\pm\left[\frac{5}{2}g_1^{(1)}Z_{1,4,3}(\hat{\bm r})-70g_1^{(3)}Z_{3,4,3}(\hat{\bm r})\right.
\notag\\&\quad\quad\quad\left.
-\frac{55}{4}g_1^{(5)}
\left\{
13Z_{5,4a,3}(\hat{\bm r})-3\sqrt{35}Z_{5,4b,3}(\hat{\bm r})
\right\}\right].
\end{align}
\end{subequations}
Since $\Gamma_7$ doublet has no orbital degrees of freedom, $\rho_{\rm e}(\hat{\bm r};\Gamma_7\pm)$ is expressed only with the even-rank $\Gamma_{1}$ harmonics.
On the other hand, $\rho_{\rm m}(\hat{\bm r};\Gamma_7\pm)$ consists of the odd-rank $\Gamma_{4}$ harmonics, which belong to the same irreducible representation of the magnetic dipole moment.
Taking an average over the time-reversal pair, we have no magnetic charge density for $\Gamma_7$ states.
$\Gamma_7$ wave functions are shown in Fig. \ref{wf_ce}(a).

\begin{figure}[tb]
\begin{center}
\includegraphics[width=8.7cm]{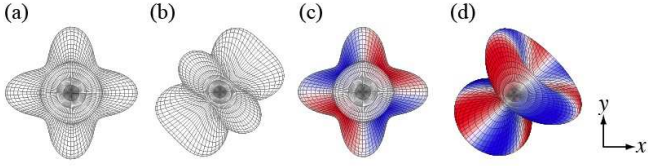}
\end{center}
\caption{(Color online) The charge and the magnetic charge densities of $\Gamma_8$ in the ordered phase, (a) disorder, (b) $\Gamma_{5g3}$, (c) $\Gamma_{2u}$ and (d) $\Gamma_{5u}^{[111]}$.}
\label{wf_order}
\end{figure}

Similarly, $\Gamma_8$ wave functions are shown in Fig. \ref{wf_ce}(b) and (c).
Taking an average over $\Gamma_8$ quartet, we have the full-symmetric charge density,
\begin{equation}
\rho_{\rm e}(\hat{\bm r};\Gamma_8)\equiv\frac{1}{4}\sum_\gamma\rho_{\rm e}(\hat{\bm r};\Gamma_8\gamma)=g_1^{(0)}+\frac{45}{2}\sqrt{21}g_1^{(4)}Z_{4,1,1}(\hat{\bm r}),
\end{equation}
and the magnetic charge density vanishes.
The averaged wave function is shown in Fig. \ref{wf_order}(a).

The high-temperature phase II is considered as the antiferro (AF) $\Gamma_{5g}$ quadrupole order\cite{Nakao01}.
Taking $z$ axis as a quantization axis, $\Gamma_{5g3}$ molecular field lifts the quartet into two doublets.
The lower doublet is expressed as
\begin{equation}
\Ket{5g3;\pm}=\frac{1}{\sqrt{2}}\left[\pm \Ket{\Gamma_8;a\pm}-i\Ket{\Gamma_8;b\pm}\right].
\end{equation}
An average over the doublet states yields the charge density,
\begin{align}
&\rho_{\rm e}(\hat{\bm r};5g3)\equiv\frac{1}{2}\sum_\gamma\rho_{\rm e}(\hat{\bm r};5g3\gamma)=\rho_{\rm e}(\hat{\bm r};\Gamma_8)
\notag\\&\quad\quad\quad\quad
-5g_1^{(2)}Z_{2,5,3}(\hat{\bm r})-45\sqrt{15}g_1^{(4)}Z_{4,5,3}(\hat{\bm r}),
\end{align}
while the magnetic charge density vanishes.
It is natural that the deviation from the cubic symmetry is characterized by the even-rank $\Gamma_{5}$ $\gamma=3$ harmonics.
The averaged wave function is shown in Fig. \ref{wf_order}(b).

In Table \ref{symconst}, there is the third-order coupling among $2u$-$4Au3$-$5g3$ multipoles in the GL free energy.
When we apply a magnetic field in [001], the uniform magnetic dipole moment ($\Gamma_{4Au3}$) arises.
As a result, the AF $\Gamma_{2u}$ octupole is induced through the 3rd-order coupling. 
Thus, the phase II is more stabilized as the magnetic field increases, provided that the AF $2u$-$2u$ exchange coupling presents.
Similarly, a magnetic field in [110] with the AF $\Gamma_{5g1}+\Gamma_{5g2}$ order induces the AF magnetic dipole moment ($\Gamma_{4Au3}$), which was observed by the neutron scattering.
This mechanism was discussed by Shiina et al. in the mean-field approximation\cite{Shiina97}, which first indicated that the magnetic octupole plays an important role behind anomalous phenomena\cite{Sakai97}.
The low-temperature phase III\cite{Effantin85} was examined similarly by the extensive use of the GL expansion accompanied with $\Gamma_{5g}$ AF quadrupole ordering\cite{Kusunose01,Sera99}.

We briefly mention the pure $\Gamma_{2u}$ octupole order\cite{Kuramoto00}, which has not been observed so far.
$\Gamma_{2u}$ molecular field also lifts the quartet into two doublets.
The lower doublet is given by 
\begin{equation}
\Ket{2u;\pm}=\frac{1}{\sqrt{2}}\left[ \Ket{\Gamma_8;a\pm}-i\Ket{\Gamma_8;b\pm}\right].
\end{equation}
This state breaks the time-reversal symmetry, however, the uniform magnetic susceptibility remains increasing with decrease of temperature in contrast to the ordinary cusp-like behavior of a magnetic order, since there still exist two-fold degeneracy.
We have the averaged charge and magnetic charge densities,
\begin{subequations}
\begin{align}
&\rho_{\rm e}(\hat{\bm r};2u)\equiv\frac{1}{2}\sum_\gamma\rho_{\rm e}(\hat{\bm r};2u\gamma)=\rho_{\rm e}(\hat{\bm r};\Gamma_8)
\\
&\rho_{\rm m}(\hat{\bm r};2u)\equiv\frac{1}{2}\sum_\gamma\rho_{\rm m}(\hat{\bm r};2u\gamma)=-\frac{63}{2}\sqrt{5}g_1^{(3)}Z_{3,2,1}(\hat{\bm r}).
\end{align}
\end{subequations}
The charge density has the full crystal symmetry, while the magnetic charge density is characterized by the odd-rank $\Gamma_{2}$ harmonics.
The wave function is shown in Fig. \ref{wf_order}(c).

Finally, we consider the phase IV, which is considered as the AF $\Gamma_{5u}$ magnetic octupole phase\cite{Kusunose01,Kubo04,Mannix05,Kusunose05,Shiina07,Kuwahara07}.
When $\Gamma_{5u}$ molecular field is applied along the high-symmetry axes, [001], [110] and [111], the maximum eigenvalue is obtained in [111].
Namely, the easy axis for $\Gamma_{5u}$ magnetic octupole is [111].
Thus, we consider AF $\Gamma_{5u}$ magnetic octupole order, in which the operator,
\begin{equation}
\hat{X}^{5u}_i\equiv
\frac{1}{\sqrt{3}}
\left(\hat{X}_i^{10}+\hat{X}_i^{11}+\hat{X}_i^{12}\right)
\end{equation}
becomes diagonal with singlet-double-singlet eigenvalues.
The non-degenerate ground state gives rise to the cusp-like behavior in the uniform magnetic susceptibility\cite{Tayama97}.
The lowest singlet is given by
\begin{align}
&
\Ket{5u}=\frac{1}{2}\left[\frac{\left(7+4\sqrt{2}i\right)^{1/4}}{\sqrt{3}}\Ket{\Gamma_8;a+}+\frac{\sqrt{2}-i}{\sqrt{3}}\Ket{\Gamma_8;a-}
\right.\notag\\&\quad\quad\quad\quad\quad\quad\left.
-(-1)^{1/4}\Ket{\Gamma_8;b+}+\Ket{\Gamma_8;b-}\right],
\end{align}
and both densities are expressed as
\begin{subequations}
\begin{align}
&\rho_{\rm e}(\hat{\bm r};5u)=\rho_{\rm e}(\hat{\bm r};\Gamma_8)-5g_1^{(2)}\overline{Z}_{2,5}(\hat{\bm r})-45\sqrt{15}g_1^{(4)}\overline{Z}_{4,5}(\hat{\bm r}),
\\
&\rho_{\rm m}(\hat{\bm r};5u)=-21\sqrt{10}g_1^{(3)}\overline{Z}_{3,5}(\hat{\bm r})-\frac{165}{4}\sqrt{70}g_1^{(5)}\overline{Z}_{5,5}(\hat{\bm r}),
\end{align}
\end{subequations}
where
\begin{equation}
\overline{Z}_{p,\Gamma}(\hat{\bm r})=\frac{1}{\sqrt{3}}\left[ Z_{p,\Gamma,1}(\hat{\bm r})+Z_{p,\Gamma,2}(\hat{\bm r})+Z_{p,\Gamma,3}(\hat{\bm r})\right].
\end{equation}
The deviations from the cubic symmetry are characterized by the even-rank $\Gamma_{5}$ harmonics and the odd-rank $\Gamma_{5}$ harmonics, respectively.
The wave function is shown in Fig. \ref{wf_order}(d).

Let us denote the primary order parameter, $\phi\equiv \overline{X}_{5u}({\bm Q})=[X_{5u1}({\bm Q})+X_{5u2}({\bm Q})+X_{5u3}({\bm Q})]/\sqrt{3}$.
There is the coupling between $\Gamma_{5u}$ magnetic octupole and the $\Gamma_{5g}$ electric quadrupole with the principal axis [111].
We denote the secondary order parameter as $\xi\equiv \overline{X}_{5g}({\bm 0})$.
We only consider the nearest neighbor AF couplings $D^{\alpha\alpha}_{\langle i,j\rangle}=-J_\alpha<0$ for $5u$-$5u$ and $5g$-$5g$ multipoles.
Then, the relevant GL free energy is given by
\begin{equation}
{\cal F}_{\rm GL}=\frac{\alpha}{2}(T-T_\phi)\phi^2+\frac{b}{4}\phi^4+\frac{a}{2}\xi^2+c\phi^2\xi+\cdots,
\end{equation}
where we have defined the critical temperatures, $T_\alpha=6J_\alpha$.
We introduce parameters, $\alpha$, $a$, $b$ and $c$, which are evaluated from the symmetric structure constants as
$\alpha=1$, $b\sim 2T_\phi/3$, $a\sim T_\phi+T_\xi$, $c\sim T_\phi/2$.
Minimizing the free energy, we have $T$ dependences of the primary and the secondary order parameters as
\begin{align}
\phi(T)=\sqrt{A(T_\phi-T)},
\quad
\xi(T)=-\frac{c}{a}\left[\phi(T)\right]^2,
\end{align}
where $A=a\alpha/(ab-2c^2)>0$.
The primary order parameter has the ordinary square-root $T$ dependence, while the secondary has $T$-linear dependence\cite{Mannix05}.
The induced uniform order $\xi$ gives rise to the lattice distortion in [111], which is indeed observed experimentally\cite{Akatsu03}.

The contribution from the fluctuations to the free energy is
\begin{equation}
\delta{\cal F}_{\rm GL}=\frac{1}{2}
\begin{pmatrix}
\delta\phi \\ \delta\xi
\end{pmatrix}^t
\hat{\chi}^{-1}
\begin{pmatrix}
\delta\phi \\ \delta\xi
\end{pmatrix},
\end{equation}
with the inverse matrix of the susceptibility,
\begin{equation}
\hat{\chi}^{-1}=
\begin{pmatrix}
\alpha(T-T_\phi)+3b\phi^2+2c\xi & 2c\phi \\
2c\phi & a
\end{pmatrix}.
\end{equation}
Then, the susceptibilities in the disorder phase are given by
\begin{equation}
\chi_\phi(T)=\frac{1}{\alpha(T-T_\phi)},
\quad
\chi_\xi(T)=\frac{1}{a},
\quad\quad
(T>T_\phi).
\end{equation}
In the ordered phase, we obtain the susceptibilities as
\begin{equation}
\chi_\phi(T)=\frac{1}{2\alpha(T_\phi-T)},
\quad
\chi_\xi(T)=\frac{bA}{a\alpha},
\quad\quad
(T<T_\phi).
\end{equation}
The susceptibility of the primary order parameter is divergent toward $T_\phi$.
On the other hand, $\chi_\xi$ of the secondary order parameter has a discontinuity at $T_\phi$, since the correlation length for $\xi$ remains finite.
The discontinuity is given by
\begin{equation}
\Delta\chi_\xi=\chi_\xi(T_\phi-)-\chi_\xi(T_\phi+)=\frac{2c^2}{\alpha a^2}A>0.
\end{equation}
Note that $\Delta\chi_\xi$ vanishes when $c\to0$.
The change of the elastic constant in $C_{44}$ mode corresponds to $-\chi_\xi$.
Therefore, the positive jump of $\chi_\xi$ leads to a sudden softening of $C_{44}$ mode\cite{Suzuki98}.

The magnetic octupole order yields the internal magnetic fields around the Ce ion\cite{Magishi02,Takagiwa02}.
From (\ref{mag_field}), we obtain the internal magnetic field\cite{Kubo04a},
\begin{equation}
{\bm B}({\bm r})=-\frac{\sqrt{16\pi}}{r^5}\overline{M}_{3,5u}{\bm Y}_{3,5u}^4(\hat{\bm r})-\frac{\sqrt{24\pi}}{r^7}\overline{M}_{5,5u}{\bm Y}_{5,5u}^6(\hat{\bm r}),
\end{equation}
where the magnetic multipole moments are expressed from (\ref{red_def}) and (\ref{mmul5u}) as
\begin{subequations}
\begin{align}
&\overline{M}_{3,5u}=\mu_{\rm B}\Braket{r^2}g_1^{(3)}\cdot 3\sqrt{5}\phi,
\\
&\overline{M}_{5,5u}=\mu_{\rm B}\Braket{r^4}g_1^{(5)}\cdot \frac{15\sqrt{35}}{4}\phi,
\end{align}
\end{subequations}
and the relevant vector spherical harmonics are given by
\begin{subequations}
\begin{align}
&{\bm Y}_{3,5u}^4(\hat{\bm r})=
\frac{1}{\sqrt{2}\sqrt{16\pi}}
\left[
\left(Z_{41}^{\rm (c)}-\sqrt{\frac{5}{2}}Z_{40}\right){\bm e}_x-\left(Z_{41}^{\rm (s)}-\sqrt{\frac{5}{2}}Z_{40}\right){\bm e}_y
\right.\notag\\&\quad\quad\quad\quad\quad\quad\quad\quad\quad\quad\left.
+\frac{5}{2}\left(Z_{41}^{\rm (c)}-Z_{41}^{\rm (s)}\right){\bm e}_z\right],
\\
&{\bm Y}_{5,5u}^6(\hat{\bm r})=
\frac{\sqrt{5}}{\sqrt{3}\sqrt{24\pi}}
\left[
\left(Z_{61}^{\rm (c)}+\frac{\sqrt{21}}{4}Z_{60}\right){\bm e}_x
-\left(Z_{61}^{\rm (s)}+\frac{\sqrt{21}}{4}Z_{60}\right){\bm e}_y
\right.\notag\\&\quad\quad\quad\quad\quad\quad\quad\quad\quad\quad\left.
-\frac{7}{4}\left(Z_{61}^{\rm (c)}-Z_{61}^{\rm (s)}\right){\bm e}_z\right].
\end{align}
\end{subequations}
The observed thermodynamic anomalies in phase IV can be understood in this way.
The detailed mean-field analysis\cite{Kubo04} is consistent with the present discussions based on the GL expansion.

\section{Summary}
We have discussed the description of multipole degrees of freedom in the consecutive fashion.
In the restricted CEF states with the degeneracy $d$, we express the exchange model (\ref{exchange_model}) in terms of the $d^2-1$ independent operators $\hat{X}_i^\alpha$.
With extensive use of the generalized Stevens' multiplicative factors $g_n^{(p)}$, we visualize the wave functions using the formula (\ref{rhoem}).

The static RPA susceptibility for the exchange model (\ref{rpa_sus}) is useful to determine the second-order phase transition line from the disorder phase under uniform external fields.
The transition temperature is determined by (\ref{ev_eq}).
The GL free-energy expansion without the external fields (\ref{gl_fe}) describes the entanglement of the multipoles in the ordered phase.
The static RPA susceptibility in the ordered phase is given by (\ref{sus_order}).

The physical multipole moments $Q_{pq}$ and $M_{pq}$ are evaluated by (\ref{mul_val}), in which the spherical tensor operators $\hat{J}_{pq}$ are expressed as linear-combinations of the operators $\hat{X}_i^\alpha$.
The existence of the multipole moments give rise to the electric and the magnetic fields near the magnetic ions, which are determined by (\ref{ele_field}) and (\ref{mag_field}) with use of the vector spherical harmonics (\ref{def_vsh}).
The explicit example using Ce$_x$La$_{1-x}$B$_6$ are given in \S6.

The analysis of the multipole exchange system tends to be complicated without systematic descriptions.
The entanglement of the multipoles plays a key role to understand anomalous responses to external fields.
The entanglement is composed concisely in the structure constants (\ref{structure_const}), which is useful to grasp a whole structure of the system.

In recent years, the experimental techniques have been developed extensively to observe semi-quantitatively higher-rank multipoles.
A quantitative analysis of a trace of the multipole moments using the electromagnetic probes could accelerate further development.

\section*{Acknowledgment}
The author would like to acknowledge stimulating discussions with Y. Kuramoto and K. Kubo.
He also acknowledges M. Yoshizawa for leading his attention to visualization of wave functions with the magnetic profile.
This work was supported by a Grant-in-Aid for Scientific Research in Priority Area ``Skutterudite" (No.18027004) of The Ministry of Education, Culture, Sports, Science and Technology, Japan.

\appendix
\section{Vector Spherical Harmonics}
In contrast to a scalar field, the spatial rotation transforms not only the position ${\bm r}$ but also the direction of the vector field.
The uniform vector is transformed as if it has the angular momentum (``spin") $1$.
Consequently, it is natural to construct the vector spherical harmonics as a direct product of $Y_{\ell m}(\hat{\bm r})$ and the spherical unit vector of rank 1,
\begin{align}
{\bm Y}_{pq}^\ell(\hat{\bm r})&\equiv
\sum_{mm'}\Braket{\ell m;1m'|pq}Y_{\ell m}(\hat{\bm r}){\bm e}_{1m'}
\notag\\&
=(-1)^{\ell+q+1}\sqrt{2p+1}\sum_{mm'}
\begin{pmatrix}
p & \ell & 1 \\ -q & m & m'
\end{pmatrix}
Y_{\ell m}(\hat{\bm r}){\bm e}_{1m'},
\label{def_vsh}
\end{align}
where $\ell=p$, $p\pm1$.
It is easy to see by definition that ${\bm Y}_{pq}^\ell(\hat{\bm r})$ transforms like $Y_{pq}(\hat{\bm r})$ under spatial rotation.
Note that ${\bm Y}_{pq}^\ell(\hat{\bm r})$ is also an eigenfunction of the orbital angular momentum,
\begin{equation}
{\bm \ell}^2{\bm Y}_{pq}^\ell(\hat{\bm r})=\ell(\ell+1){\bm Y}_{pq}^\ell(\hat{\bm r}).
\label{orb_es}
\end{equation}
If we introduce a ``total" angular momentum from the orbital and the ``spin" angular momenta, ${\bm Y}_{pq}^\ell(\hat{\bm r})$ is its eigenfunction, and the indices $p$ and $q$ represent the quantum numbers of the magnitude and the projection of the ``total" angular momentum, respectively.

The spherical unit vector is given by the cartesian unit vectors,
\begin{align}
\begin{split}
&{\bm e}_{11}=-\frac{1}{\sqrt{2}}({\bm e}_x+i{\bm e}_y),
\\
&{\bm e}_{10}={\bm e}_z
\\
&{\bm e}_{1-1}=\frac{1}{\sqrt{2}}({\bm e}_x-i{\bm e}_y),
\end{split}
\label{sp_unit_vector}
\end{align}
which satisfy the orthogonality, ${\bm e}_{1m'}^*\cdot{\bm e}_{1m}=\delta_{mm'}$.
This definition is compatible with the real representation in \S 2.4.
An arbitrary vector is expressed as
\begin{equation}
{\bm A}=\sum_mA_{1m}{\bm e}_{1m}^*=\sum_mA_{1m}^*{\bm e}_{1m},
\end{equation}
where the spherical components of ${\bm A}$ are deduced from their cartesian components as similar to (\ref{sp_unit_vector}).
The simplest cases of the vector spherical harmonics are
\begin{equation}
{\bm Y}_{00}^0(\hat{\bm r})=0,
\quad
{\bm Y}_{1q}^0(\hat{\bm r})=\frac{1}{\sqrt{4\pi}}{\bm e}_{1q},
\quad
{\bm Y}_{00}^1(\hat{\bm r})=-\frac{1}{\sqrt{4\pi}}\hat{\bm r}.
\end{equation}
The complex conjugation is given by
\begin{equation}
\left[{\bm Y}_{pq}^\ell(\hat{\bm r})\right]^*=(-1)^{p+q+\ell+1}{\bm Y}_{p-q}^\ell(\hat{\bm r}).
\end{equation}

The vector spherical harmonics are also expressed in terms of $Y_{pq}(\hat{\bm r})$, $\hat{\bm r}$ and ${\bm\ell}$,
\begin{subequations}
\begin{align}
&{\bm Y}_{pq}^p(\hat{\bm r})=\frac{1}{\sqrt{p(p+1)}}{\bm\ell}Y_{pq}(\hat{\bm r}),
\\
&{\bm Y}_{pq}^{p-1}(\hat{\bm r})=\frac{1}{\sqrt{p(2p+1)}}\left(p\hat{\bm r}-i\hat{\bm r}\times{\bm\ell}\right)Y_{pq}(\hat{\bm r}),
\\
&{\bm Y}_{pq}^{p+1}(\hat{\bm r})=\frac{-1}{\sqrt{(p+1)(2p+1)}}\left[
(p+1)\hat{\bm r}+i\hat{\bm r}\times{\bm\ell}
\right]Y_{pq}(\hat{\bm r}).
\end{align}
\end{subequations}
These expressions are derived from the definition, (\ref{def_vsh}).

\begin{figure}[tb]
\begin{center}
\includegraphics[width=3.5cm]{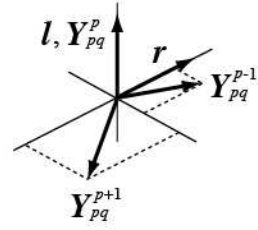}
\end{center}
\caption{The schematic relations between ${\bm Y}_{pq}^\ell$, ${\bm\ell}$ and ${\bm r}$.}
\label{direct_vsh}
\end{figure}

To characterize the direction of ${\bm Y}_{pq}^\ell(\hat{\bm r})$, the scalar and the vector products with $\hat{\bm r}$ are useful.
Using identities, $\hat{\bm r}\cdot{\bm\ell}=\hat{\bm r}\cdot(\hat{\bm r}\times{\bm\ell})=0$ and $\hat{\bm r}\times(\hat{\bm r}\times{\bm\ell})=-{\bm\ell}$, we obtain
\begin{subequations}
\begin{align}
&\hat{\bm r}\cdot{\bm Y}_{pq}^p(\hat{\bm r})=0,
\\
&\hat{\bm r}\cdot{\bm Y}_{pq}^{p-1}(\hat{\bm r})=\sqrt{\frac{p}{2p+1}}Y_{pq}(\hat{\bm r}),
\\
&\hat{\bm r}\cdot{\bm Y}_{pq}^{p+1}(\hat{\bm r})=-\sqrt{\frac{p+1}{2p+1}}Y_{pq}(\hat{\bm r}),
\end{align}
\end{subequations}
and
\begin{subequations}
\begin{align}
&-i\hat{\bm r}\times{\bm Y}_{pq}^p(\hat{\bm r})=\sqrt{\frac{p+1}{2p+1}}{\bm Y}_{pq}^{p-1}(\hat{\bm r})+\sqrt{\frac{p}{2p+1}}{\bm Y}_{pq}^{p+1}(\hat{\bm r}),
\\
&-i\hat{\bm r}\times{\bm Y}_{pq}^{p-1}(\hat{\bm r})=\sqrt{\frac{p+1}{2p+1}}{\bm Y}_{pq}^p(\hat{\bm r}),
\label{vec_prod}
\\
&-i\hat{\bm r}\times{\bm Y}_{pq}^{p+1}(\hat{\bm r})=\sqrt{\frac{p}{2p+1}}{\bm Y}^p_{pq}(\hat{\bm r}).
\end{align}
\end{subequations}
The schematic relations between ${\bm Y}_{pq}^\ell$, ${\bm\ell}$ and ${\bm r}$ are shown in Fig. \ref{direct_vsh}.

Since the differential operator ${\bm\nabla}$ also has a vector property, derivatives can be expanded in terms of ${\bm Y}_{pq}^\ell(\hat{\bm r})$.
With the help of the identity, ${\bm\nabla}=\hat{\bm r}(\partial/\partial r)-(i/r)(\hat{\bm r}\times{\bm\ell})$, we obtain
\begin{gather}
{\bm\nabla}\left(f(r)Y_{pq}(\hat{\bm r})\right)=
{\bm Y}_{pq}^{p-1}(\hat{\bm r})\sqrt{\frac{p}{2p+1}}\left(
\frac{p+1}{r}+\frac{d}{dr}
\right)f(r)
\notag\\
+{\bm Y}_{pq}^{p+1}(\hat{\bm r})\sqrt{\frac{p+1}{2p+1}}\left(
\frac{p}{r}-\frac{d}{dr}
\right)f(r),
\end{gather}
where $f(r)$ is an arbitrary function of the radial coordinate $r$.

The divergence and the rotation are also given by
\begin{subequations}
\begin{align}
&-{\bm\nabla}\cdot\left(f(r){\bm Y}_{pq}^{p}(\hat{\bm r})\right)=0,
\\
&-{\bm\nabla}\cdot\left(f(r){\bm Y}_{pq}^{p-1}(\hat{\bm r})\right)=Y_{pq}(\hat{\bm r})\sqrt{\frac{p}{2p+1}}\left(\frac{p-1}{r}-\frac{d}{dr}\right)f(r),
\\
&-{\bm\nabla}\cdot\left(f(r){\bm Y}_{pq}^{p+1}(\hat{\bm r})\right)=Y_{pq}(\hat{\bm r})\sqrt{\frac{p+1}{2p+1}}\left(\frac{p+2}{r}+\frac{d}{dr}\right)f(r).
\end{align}
\label{div_vsh}
\end{subequations}
\begin{subequations}
\begin{align}
&-i{\bm\nabla}\times\left(f(r){\bm Y}_{pq}^{p}(\hat{\bm r})\right)={\bm Y}_{pq}^{p-1}(\hat{\bm r})\sqrt{\frac{p+1}{2p+1}}\left(\frac{p+1}{r}+\frac{d}{dr}\right)f(r)
\notag\\&\quad\quad\quad
-{\bm Y}_{pq}^{p+1}(\hat{\bm r})\sqrt{\frac{p}{2p+1}}\left(\frac{p}{r}-\frac{d}{dr}\right)f(r),
\\
&-i{\bm\nabla}\times\left(f(r){\bm Y}_{pq}^{p-1}(\hat{\bm r})\right)=-{\bm Y}_{pq}^{p}(\hat{\bm r})\sqrt{\frac{p+1}{2p+1}}\left(\frac{p-1}{r}-\frac{d}{dr}\right)f(r),
\\
&-i{\bm\nabla}\times\left(f(r){\bm Y}_{pq}^{p+1}(\hat{\bm r})\right)=
{\bm Y}_{pq}^{p}(\hat{\bm r})\sqrt{\frac{p}{2p+1}}\left(\frac{p+2}{r}+\frac{d}{dr}\right)f(r).
\end{align}
\end{subequations}
From those formula, we have some useful relations,
\begin{subequations}
\begin{align}
&{\bm\nabla}\cdot\left(f{\bm Y}_{pq}^p\right)={\bm\nabla}\cdot\left(r^{p-1}{\bm Y}_{pq}^p\right)=
{\bm\nabla}\cdot\left(\frac{{\bm Y}_{pq}^{p+1}}{r^{p+2}}\right)=0,
\label{div0}
\\
&{\bm\nabla}\times\left(r^{p-1}{\bm Y}_{pq}^{p-1}\right)={\bm\nabla}\times\left(\frac{{\bm Y}_{pq}^{p+1}}{r^{p+2}}\right)=0,
\\
&{\bm\nabla}\left(r^p Y_{pq}\right)=-\sqrt{\frac{p}{p+1}}{\bm\nabla}\times\left(ir^p{\bm Y}_{pq}^p\right)=r^{p-1}\sqrt{p(2p+1)}{\bm Y}_{pq}^{p-1},
\label{grad_sh}
\\
&{\bm\nabla}\left(\frac{Y_{pq}}{r^{p+1}}\right)=
-\sqrt{\frac{p+1}{p}}{\bm\nabla}\times\left(\frac{{\bm Y}_{pq}^p}{ir^{p+1}}\right)
=
\frac{\sqrt{(p+1)(2p+1)}}{r^{p+2}}{\bm Y}_{pq}^{p+1}.
\label{grad_sh2}
\end{align}
\end{subequations}

A scalar product of two vector spherical harmonics are expanded in terms of $Y_{\ell m}(\hat{\bm r})$ as
\begin{align}
&{\bm Y}_{p_1q_1}^{\ell_1*}(\hat{\bm r})\cdot{\bm Y}_{p_2q_2}^{\ell_2}(\hat{\bm r})
=
\notag\\
&\quad\quad
(-1)^{q_2+1}\sqrt{\frac{(2p_1+1)(2p_2+1)(2\ell_1+1)(2\ell_2+1)}{4\pi}}
\notag\\
&\quad\quad\quad\quad\times
\sum_{\ell m}
\sqrt{2\ell+1}
\begin{pmatrix}
\ell & \ell_1 & \ell_2 \\ 0 & 0 & 0
\end{pmatrix}
\begin{pmatrix}
\ell & p_1 & p_2 \\ -m & -q_1 & q_2
\end{pmatrix}
\notag\\
&\quad\quad\quad\quad\quad\quad\times
\begin{Bmatrix}
\ell_1 & \ell_2 & \ell \\ p_2 & p_1 & 1
\end{Bmatrix}
Y_{\ell m}(\hat{\bm r}),
\end{align}
where curly bracket is the $6j$ symbol\cite{Landau81}.
Using this formula, we evaluate the angular integral,
\begin{align}
&
\int d\hat{\bm r}Y_{\ell m}^*(\hat{\bm r})
\left[
{\bm Y}_{p_1q_1}^{\ell_1*}(\hat{\bm r})\cdot{\bm Y}_{p_2q_2}^{\ell_2}(\hat{\bm r})
\right]
=
\notag\\&\quad\quad
(-1)^{q_2+1}\sqrt{\frac{(2p_1+1)(2p_2+1)(2\ell_1+1)(2\ell_2+1)(2\ell+1)}{4\pi}}
\notag\\
&\quad\quad\quad\quad\times
\begin{pmatrix}
\ell & \ell_1 & \ell_2 \\ 0 & 0 & 0
\end{pmatrix}
\begin{pmatrix}
\ell & p_1 & p_2 \\ -m & -q_1 & q_2
\end{pmatrix}
\begin{Bmatrix}
\ell_1 & \ell_2 & \ell \\ p_2 & p_1 & 1
\end{Bmatrix}.
\label{vsh_formula1}
\end{align}
In the case of $\ell=m=0$, we have the projective orthogonality relation,
\begin{equation}
\int d\hat{\bm r}\,\,
{\bm Y}_{p_1q_1}^{\ell_1*}(\hat{\bm r})\cdot{\bm Y}_{p_2q_2}^{\ell_2}(\hat{\bm r})
=\delta_{p_1p_2}\delta_{q_1q_2}\delta_{\ell_1\ell_2}.
\label{ortho_vsh}
\end{equation}

\section{Details of Multipole Expansion}

\subsection{The scalar potential}
We write the solution of the Poisson equation in the form,
\begin{equation}
\phi({\bm r})=\sum_{p=0}^\infty \sum_{q=-p}^p Z_{pq}(\hat{\bm r})\theta_{pq}(r).
\end{equation}
Substituting this solution into the Poisson equation, we have
\begin{equation}
\sum_{pq}Z_{pq}(\hat{\bm r})\left[{\bm\nabla}^2_r-\frac{p(p+1)}{r^2}\right]\theta_{pq}(r)=-4\pi\rho({\bm r}),
\end{equation}
where ${\bm\nabla}^2_r$ is the radial part of the laplacian.
Using the orthogonality of the spherical harmonics, we obtain the differential equation for $\theta_{pq}(r)$,
\begin{equation}
\left[{\bm\nabla}^2_r-\frac{p(p+1)}{r^2}\right]\theta_{pq}(r)=-(2p+1)\int d\hat{\bm r}\,Z_{pq}^*(\hat{\bm r})\rho({\bm r}).
\label{theta_eq}
\end{equation}

In order to solve this equation, we consider the radial part of the inhomogeneous Helmholtz equation with a point source,
\begin{equation}
\left[
{\bm\nabla}^2_r+k^2-\frac{p(p+1)}{r^2}\right]g(r,r')=-\frac{\delta(r-r')}{r^2}.
\end{equation}
The solution (the Green's function) is given by
\begin{equation}
g(r,r')=ikj_p(kr_<)h_p^{(1)}(kr_>),
\end{equation}
where $j_p(x)$ and $h_p^{(1)}(x)$ are the spherical Bessel function and the spherical Hankel function of the first kind, respectively, and $r_<=\min(r,r')$, $r_>=\max(r,r')$.
In the limit of $k\to0$, the Green's function becomes
\begin{equation}
g(r,r')=\frac{1}{2p+1}\frac{r_<^p}{r_>^{p+1}}.
\label{gf}
\end{equation}
Using the Green's function, we have the solution for (\ref{theta_eq})
\begin{equation}
\theta_{pq}(r)=\int d{\bm r}'\frac{r_<^p}{r_>^{p+1}}Z_{pq}^*(\hat{\bm r}')\rho({\bm r}').
\end{equation}
For regions outside the source distribution, we have $r_<=r'$ and $r_>=r$.
Then, the multipole expansion for the scalar field is given by
\begin{equation}
\phi({\bm r})=\sum_{p=0}^\infty \sum_{q=-p}^p \frac{1}{r^{p+1}}Z_{pq}(\hat{\bm r})Q_{pq},
\end{equation}
with the electric multipole moment,
\begin{equation}
Q_{pq}=\int d{\bm r}\, r^pZ_{pq}^*(\hat{\bm r})\rho({\bm r}).
\label{e_multipole}
\end{equation}

\subsection{The vector potential}
The derivation here essentially follows that given by Schwartz in the context of the hyperfine structure of nuclear matter.\cite{Schwartz55}.
In the case of the gauge ${\bm\nabla}\cdot{\bm A}=0$, the vector potential is parallel to ${\bm Y}_{pq}^p(\hat{\bm r})$ due to (\ref{div_vsh}).
Namely, we express the solution in the form,
\begin{equation}
{\bm A}({\bm r})=\sum_{p=0}^\infty \sum_{q=-p}^p {\bm Z}_{pq}(\hat{\bm r})\zeta_{pq}(r),
\end{equation}
where we have introduced
\begin{equation}
{\bm Z}_{pq}(\hat{\bm r})=-i\sqrt{\frac{4\pi (p+1)}{p(2p+1)}}{\bm Y}_{pq}^p(\hat{\bm r})
=\frac{{\bm\ell}Z_{pq}(\hat{\bm r})}{ip},
\end{equation}
for notational simplicity.
Substituting this into the Poisson equation and using (\ref{orb_es}), we have
\begin{equation}
\sum_{pq}{\bm Z}_{pq}(\hat{\bm r})\left[{\bm\nabla}^2_r-\frac{p(p+1)}{r^2}\right]\zeta_{pq}(r)=-\frac{4\pi}{c}{\bm j}({\bm r}).
\end{equation}
By the orthogonality, (\ref{ortho_vsh}), we have the differential equation for $\zeta_{pq}(r)$,
\begin{equation}
\left[{\bm\nabla}^2_r-\frac{p(p+1)}{r^2}\right]\zeta_{pq}(r)=-\frac{p(2p+1)}{c(p+1)}\int d\hat{\bm r} {\bm Z}_{pq}^*(\hat{\bm r})\cdot{\bm j}({\bm r}).
\label{eq_zeta}
\end{equation}
We obtain the solution using the Green's function (\ref{gf}),
\begin{equation}
\zeta_{pq}(r)=\frac{p}{c(p+1)}\int d{\bm r}'\frac{r_<^p}{r_>^{p+1}}{\bm Z}_{pq}^*(\hat{\bm r}')\cdot{\bm j}({\bm r}').
\end{equation}
For regions outside the source distribution, we have
\begin{equation}
{\bm A}({\bm r})=\sum_{p=0}^\infty \sum_{q=-p}^p \frac{1}{r^{p+1}}{\bm Z}_{pq}(\hat{\bm r})M_{pq},
\end{equation}
where we have defined the magnetic multipole moment as
\begin{equation}
M_{pq}=\frac{p}{c(p+1)}\int d{\bm r}\, r^p\left[{\bm Z}_{pq}^*(\hat{\bm r})\cdot{\bm j}({\bm r})\right].
\end{equation}
Using (\ref{vec_prod}), we rewrite the magnetic multipole moment as
\begin{align}
M_{pq}&
=\sqrt{4\pi p}\int d{\bm r}\, r^{p-1}{\bm Y}^{p-1*}_{pq}(\hat{\bm r})\cdot\frac{{\bm r}\times{\bm j}({\bm r})}{c(p+1)}.
\label{m_multipole1}
\end{align}

Now, we introduce the magnetization density by
\begin{equation}
{\bm j}({\bm r})=c{\bm\nabla}\times{\bm M}({\bm r}).
\end{equation}
With use of the identity,
\begin{gather}
\frac{{\bm r}\times{\bm j}}{c}={\bm r}\times({\bm\nabla}\times{\bm M})={\bm\nabla}\left({\bm r}\cdot{\bm M}\right)-\left[1+({\bm r}\cdot{\bm\nabla})\right]{\bm M},
\end{gather}
and the integrations by part, we obtain
\begin{align}
M_{pq}&=\sqrt{4\pi p}\int d{\bm r}\left[\left(2+r\frac{\partial}{\partial r}\right)r^{p-1}{\bm Y}_{pq}^{p-1*}(\hat{\bm r}) \right]\cdot\frac{{\bm M}({\bm r})}{p+1}
\notag\\
&=\sqrt{4\pi p}\int d{\bm r}\,r^{p-1}{\bm Y}_{pq}^{p-1*}(\hat{\bm r})\cdot{\bm M}({\bm r})
\notag\\
&=\int d{\bm r}\,{\bm\nabla}\left[r^{p}Z_{pq}^*(\hat{\bm r})\right]\cdot{\bm M}({\bm r}),
\label{m_multipole2}
\end{align}
where we have used (\ref{div0}) and (\ref{grad_sh}).

The current density is expressed in terms of the angular-momentum density.
The orbital current has the relation
\begin{equation}
{\bm r}\times{\bm j}_{\rm orb.}({\bm r})=2\mu_{\rm B}c{\bm\ell}({\bm r}),
\end{equation}
while the spin current is ${\bm j}_{\rm spin}=2\mu_{\rm B}c{\bm\nabla}\times{\bm s}({\bm r})$ by definition.
The latter gives the spin magnetization density as ${\bm M}_{\rm spin}({\bm r})=2\mu_{\rm B}{\bm s}({\bm r})$.
Using (\ref{m_multipole1}) and (\ref{m_multipole2}) for the orbital and the spin parts, respectively, we obtain
\begin{equation}
M_{pq}=\mu_{\rm B}\int d{\bm r}\,{\bm\nabla}\left[r^{p}Z_{pq}^*(\hat{\bm r})\right]\cdot
\left[\frac{2{\bm\ell}({\bm r})}{p+1}+2{\bm s}({\bm r})\right].
\end{equation}
The orbital angular-momentum densities are expressed in terms of the $f$-electron operators as
\begin{equation}
{\bm \ell}({\bm r})=\Braket{\sum_j\delta({\bm r}-{\bm r}_j){\bm\ell}_j}_f,
\quad
{\bm s}({\bm r})=\Braket{\sum_j\delta({\bm r}-{\bm r}_j){\bm s}_j}_f.
\end{equation}

\section{Derivation of Generalized Stevens' factors}
Let us express the wave function of the multiplet, $^{J}L_{2S+1}$, with the $f^n$ configuration in the Russell-Sanders scheme as
\begin{equation}
\Ket{n JM}=(-1)^{L-S+M}\sqrt{2J+1}\sum_{m\sigma}
\begin{pmatrix}
J & L & S \\ -M & m & \sigma
\end{pmatrix}
\Ket{n Lm}\Ket{n S\sigma}.
\end{equation}
We first express the reduced matrix elements of the multipole operators in terms of the expectation value of the particular orbital state\cite{Sievers82}, $\Ket{nLL}$.
Then, we derive the expectation value for the Hund's-rule ground state with the maximum $L$.

\subsection{The reduced matrix elements of multipole operators}
For an operator of rank $p$ which acts only on the orbital part of the wave function, we have the relation\cite{Inui96},
\begin{align}
&\Braket{n J||f_p(L)||n J}=
\notag\\&\quad\quad
(-1)^{J+L+S+p}
(2J+1)
\begin{Bmatrix}
J & J & p \\ L & L & S
\end{Bmatrix}
\Braket{n L||f_p||n L}
\notag\\&\quad\quad
=\lambda(p,J,L,S)\Braket{nLL|f_{p0}|nLL},
\label{red_orbital}
\end{align}
where we have defined
\begin{equation}
\lambda(p,J,L,S)=
(-1)^{J+L+S+p}
(2J+1)
\frac{
\begin{Bmatrix}
J & J & p \\ L & L & S
\end{Bmatrix}
}{
\begin{pmatrix}
p & L & L \\ 0 & L & -L
\end{pmatrix}
}.
\end{equation}
Note that in the case of $J=L+S$, the coefficient becomes independent of $L$ and $S$,
\begin{equation}
\lambda(p,J,L,S)=
\begin{pmatrix}
p & J & J \\ 0 & J & -J
\end{pmatrix}^{-1},
\quad\text{(for $S=J-L$)}.
\end{equation}

With the help of (\ref{red_orbital}), the reduced matrix element of the electric multipole operator (\ref{e_mul_op}) is given by
\begin{align}
\Braket{nJ||\hat{Q}_p||nJ}&=\sum_j\Braket{nJ||\hat{Q}_p(j)||nJ}
\notag\\&=
\lambda(p,J,L,S)
n
\Braket{nLL|\hat{Q}_{p0}(j)|nLL},
\label{red_ele}
\end{align}
with
\begin{equation}
\hat{Q}_{p0}(j)=-e\int d{\bm r}\delta({\bm r}-{\bm r}_j)r^p Z_{p0}(\hat{\bm r}).
\end{equation}
Here, we have used that $\Braket{nLL|\hat{Q}_{p0}(j)|nLL}$ is independent of $j$, as will be shown shortly.

Similarly, we apply the following relation to the magnetic multipole operator, (\ref{m_mul_op}),
\begin{align}
&\Braket{n J||\left(f(r){\bm Y}_{pq}^\ell(\hat{\bm r})\cdot{\bm g}(S)\right)_p||n J}=
(2J+1)\sqrt{2p+1}
\notag\\&\quad\quad\quad\quad\quad\times
\begin{Bmatrix}
J & J & p \\ L & L & \ell \\ S & S & 1
\end{Bmatrix}
\Braket{n L||fY_\ell||n L}
\Braket{n S||g_1||n S}
\notag\\&\quad\quad
\equiv\lambda(p,J,L,S,\ell)
\frac{\sqrt{2p+1}}{\sqrt{S(S+1)(2S+1)}}
\notag\\&\quad\quad\quad\quad\quad\times
\Braket{nLL|f(r)Y_{\ell0}(\hat{\bm r})|nLL}
\Braket{n S||g_1||n S},
\end{align}
where we have defined
\begin{align}
\lambda(p,J,L,S,\ell)=(2J+1)\sqrt{S(S+1)(2S+1)}
\frac{
\begin{Bmatrix}
J & J & p \\ L & L & \ell \\ S & S & 1
\end{Bmatrix}
}{
\begin{pmatrix}
\ell & L & L \\ 0 & L & -L
\end{pmatrix}
},
\end{align}
and $3\times3$ curly bracket denotes the $9j$ symbol\cite{Landau81}.
Note that the inner product ${\bm Y}_{pq}^{\ell}(\hat{\bm r})\cdot{\bm g}$ transforms like $Y_{pq}(\hat{\bm r})$ under spatial rotation.
The result is
\begin{align}
&\Braket{nJ||\hat{M}_p||nJ}
=\sum_j\Braket{nJ||\hat{M}_p^{\rm orb.}(j)+\left[\hat{M}_{pq}^{p-1}(j)\cdot{\bm s}_j\right]_p||nJ}
\notag\\&\quad\quad
=
\lambda(p,J,L,S)
n
\Braket{nLL|\hat{M}_{p0}^{\rm orb.}(j)|nLL}
\notag\\&\quad\quad\quad\quad
+\lambda(p,J,L,S,p-1)
\Braket{nLL|\hat{f}_{p-10}(j)|nLL},
\label{red_mag}
\end{align}
with
\begin{align}
&\hat{M}_{p0}^{\rm orb.}(j)=\frac{2\mu_{\rm B}}{p+1}\int d{\bm r}\delta({\bm r}-{\bm r}_j){\bm\nabla}\left(
r^pZ_{p0}(\hat{\bm r})
\right)\cdot{\bm\ell}_j,
\\&
\hat{f}_{p-10}(j)=2\mu_{\rm B}\sqrt{p(4p^2-1)}\int d{\bm r}\delta({\bm r}-{\bm r}_j)
r^{p-1}Z_{p-10}(\hat{\bm r}).
\end{align}

\subsection{The expectation values of the Hund's-rule ground state}
The Hund's-rule ground multiplet in the Russell-Sanders scheme with $f^n$ configuration is characterized by the quantum numbers $(JLS)$,
\begin{equation}
J=|L-S|,
\,\,
L=\sum_{j=1}^n(4-j),
\,\,
S=n/2,
\end{equation}
for $n\le7$, otherwise
\begin{equation}
J=L+S,
\,\,
L=\sum_{j=1}^{n-7}(4-j),
\,\,
S=7-n/2.
\end{equation}

The orbital wave function is expressed by the Slater determinant of the one-body atomic wave function, $\varphi_{4-m}({\bm r})=R_f(r)Y_{3m}(\hat{\bm r})$,
\begin{align}
\Ket{nLL}&=\frac{1}{\sqrt{n!}}
\begin{vmatrix}
\,\,\varphi_1({\bm r}_1) & \varphi_1({\bm r}_2) & \cdots & \varphi_1({\bm r}_n)\,\, \\
\,\,\varphi_2({\bm r}_1) & \varphi_2({\bm r}_2) & \cdots & \varphi_2({\bm r}_n)\,\, \\
\,\,\vdots & \vdots & \ddots & \vdots\,\, \\
\,\,\varphi_n({\bm r}_1) & \varphi_n({\bm r}_2) & \cdots & \varphi_n({\bm r}_n)\,\, \\
\end{vmatrix}
\notag\\
&=\sum_m \frac{(-1)^{m-j}}{\sqrt{n}}\varphi_m({\bm r}_j)\Ket{n-1;mj},
\end{align}
where $\Ket{n-1;mj}$ is $(n-1)\times(n-1)$ Slater determinant from which $\varphi_m$ and ${\bm r}_j$ are eliminated.

We consider an operator in the form,
\begin{equation}
\hat{\cal O}(j)=\int d{\bm r}\delta({\bm r}-{\bm r}_j)f(r)g(\hat{\bm r}),
\end{equation}
then we evaluate the expectation value as
\begin{align}
I\left(\hat{\cal O}(j)\right)&=n\Braket{nLL|\hat{\cal O}(j)|nLL}
\notag\\&
=\Braket{f(r)}\sum_m\int d\hat{\bm r}Y_{3m}^*(\hat{\bm r})g(\hat{\bm r})Y_{3m}(\hat{\bm r}),
\end{align}
where $\Braket{f(r)}$ is the radial average, and $I(\hat{\cal O}(j))$ is independent of $j$.
Using the formula
\begin{align}
&\int d\hat{\bm r}
Y_{p_1q_1}(\hat{\bm r})Y_{p_2q_2}(\hat{\bm r})Y_{p_3q_3}(\hat{\bm r})
=
\sqrt{\frac{(2p_1+1)(2p_2+1)(2p_3+1)}{4\pi}}
\notag\\&\quad\quad\quad\quad\quad\times
\begin{pmatrix}
p_1 & p_2 & p_3 \\ 0 & 0 & 0
\end{pmatrix}
\begin{pmatrix}
p_1 & p_2 & p_3 \\ q_1 & q_2 & q_3
\end{pmatrix},
\label{selection_sh}
\end{align}
and (\ref{vsh_formula1}), we obtain the required expectation values,
\begin{subequations}
\begin{align}
&I\left(\hat{Q}_{p0}(j)\right)=-e\Braket{r^p}K_{pp}(n),
\\
&I\left(\hat{M}_{p0}^{\rm orb.}(j)\right)=-4\sqrt{21}\mu_{\rm B}\Braket{r^{p-1}}\frac{\sqrt{p(2p+1)(2p-1)}}{p+1}
\notag\\&\quad\quad\quad\quad\quad\quad\times
\begin{Bmatrix}
p-1 & p & 1 \\ 3 & 3 & 3
\end{Bmatrix}
K_{p-1,p}(n)
\notag\\&\quad\quad
=\mu_{\rm B}\Braket{r^{p-1}}(-1)^{p+1}\frac{p\sqrt{49-p^2}}{p+1}K_{p-1,p}(n),
\\
&I\left(\hat{f}_{p-10}(j)\right)=\mu_{\rm B}\Braket{r^{p-1}}2\sqrt{p(4p^2-1)}
K_{p-1,p-1}(n),
\end{align}
\label{angle_av}
\end{subequations}
where we have defined
\begin{subequations}
\begin{align}
&K_{pk}(n)=7
\begin{pmatrix}
p & 3 & 3 \\ 0 & 0 & 0
\end{pmatrix}
\sum_{j=1}^n
\begin{pmatrix}
k & 3 & 3 \\ 0 & 4-j & j-4
\end{pmatrix}
(-1)^j,
\,\,
(n\le7),
\\&
K_{pk}(n)=7
\begin{pmatrix}
p & 3 & 3 \\ 0 & 0 & 0
\end{pmatrix}
\left[
\sum_{j=1}^{n-7}
\begin{pmatrix}
k & 3 & 3 \\ 0 & 4-j & j-4
\end{pmatrix}
(-1)^j
\right.\notag\\&\quad\quad\quad\quad\quad\quad\quad\quad\quad\quad\left.
-\sqrt{7}\delta_{k0}\right],
(\text{otherwise}).
\end{align}
\end{subequations}

From (\ref{red_ele}), (\ref{red_mag}) and (\ref{angle_av}), we have the reduced matrix elements of the multipole operators
\begin{align}
&\frac{\Braket{nJ||\hat{Q}_p||nJ}}{-e\Braket{r^p}}=
\lambda(p,J,L,S)
K_{pp}(n),
\label{sf_ele}
\\
&
\frac{\Braket{nJ||\hat{M}_p||nJ}}{\mu_{\rm B}\Braket{r^{p-1}}}=
\frac{(-1)^{p+1}p}{p+1}\sqrt{49-p^2}
\lambda(p,J,L,S)
K_{p-1,p}(n)
\notag\\
&\quad
+\frac{2}{n}
\sqrt{p(4p^2-1)}
\lambda(p,J,L,S,p-1)K_{p-1,p-1}(n).
\label{sf_mag}
\end{align}

\subsection{Generalized Stevens' factors}
Applying the Wigner-Eckart theorem to (\ref{red_def}), we obtain
\begin{align}
&
g_n^{(p)}=\frac{\Braket{nJ||\hat{Q}_p||nJ}}{-e\Braket{r^p}\Braket{J||\hat{J}_p||J}},
\\&
g_n^{(p)}=\frac{\Braket{nJ||\hat{M}_p||nJ}}{\mu_{\rm B}\Braket{r^{p-1}}\Braket{J||\hat{J}_p||J}}.
\end{align}
With (\ref{sf_ele}), (\ref{sf_mag}) and (\ref{jp_red}), the generalized Stevens' multiplicative factors $g_n^{(p)}$ are evaluated in Table \ref{stevens} for the Hund's-rule ground multiplet with $f^n$ configuration, in which we also give the ratio of the orbital and the spin contributions to the magnetic multipoles,
\begin{equation}
r_n^{(p)}=\frac{g_n^{(p)}(\text{orbital})}{g_n^{(p)}(\text{spin})}.
\end{equation}
Note that the orbital contribution in $r_n^{(3)}$ vanishes for $L=5$ due to $K_{23}(n)=0$.

In the case of $p=1$, we have
\begin{align}
&K_{00}(n)=n,
\quad
K_{01}(n)=\frac{L}{2\sqrt{3}},
\notag\\&
\frac{\lambda(1,J,L,S)}{\Braket{J||\hat{J}_1||J}}=\frac{1}{2L}\left[1+\frac{L(L+1)-S(S+1)}{J(J+1)}\right],
\notag\\&
\frac{\lambda(1,J,L,S,0)}{\Braket{J||\hat{J}_1||J}}=\frac{1}{2\sqrt{3}}\left[1-\frac{L(L+1)-S(S+1)}{J(J+1)}\right],
\end{align}
and we obtain
\begin{equation}
g_n^{(1)}=\frac{3}{2}-\frac{L(L+1)-S(S+1)}{J(J+1)}.
\end{equation}
This is nothing but the Land\'e's $g$ factor.

\onecolumn
\begin{table*}[tb]
\caption{The tesseral harmonics (multiplied by $r^p$) in the cartesian coordinate. The parity is given by $(-1)^p$.}
\label{scaled_sh}
\begin{tabular}{ccll}
\hline
$p$ & $q$ & \;\;$r^pZ^{\rm (c)}_{pq}(\hat{\bm r})$\;\; or \;\;$r^pZ_{p0}(\hat{\bm r})$ & $ r^pZ^{\rm (s)}_{pq}(\hat{\bm r})$ \\ \hline
0 & 0 & 1 & \\ \hline
1 & 0 & $\displaystyle z$ & \\
  & 1 & $\displaystyle x$ & $\displaystyle y$ \\ \hline
2 & 0 & $\displaystyle\frac{1}{2}\left(3z^2-r^2\right)$ & \\
  & 1 & $\displaystyle \sqrt{3}zx$ & $\displaystyle \sqrt{3}yz$ \\
  & 2 & $\displaystyle \frac{\sqrt{3}}{2}\left(x^2-y^2\right)$ & $\displaystyle \sqrt{3}xy$ \\ \hline
3 & 0 & $\displaystyle \frac{1}{2}z\left(5z^2-3r^2\right)$ & \\
  & 1 & $\displaystyle \frac{\sqrt{6}}{4}x\left(5z^2-r^2\right)$ & $\displaystyle \frac{\sqrt{6}}{4}y\left(5z^2-r^2\right)$ \\
  & 2 & $\displaystyle \frac{\sqrt{15}}{2}z\left(x^2-y^2\right)$ & $\displaystyle \sqrt{15}xyz$ \\
  & 3 & $\displaystyle \frac{\sqrt{10}}{4}x\left(x^2-3y^2\right)$ & $\displaystyle \frac{\sqrt{10}}{4}y\left(3x^2-y^2\right)$ \\ \hline
4 & 0 & $\displaystyle \frac{1}{8}\left(35z^4-30z^2r^2+3r^4\right)$ & \\
  & 1 & $\displaystyle \frac{\sqrt{10}}{4}zx\left(7z^2-3r^2\right)$ & $\displaystyle \frac{\sqrt{10}}{4}yz\left(7z^2-3r^2\right)$ \\
  & 2 & $\displaystyle \frac{\sqrt{5}}{4}\left(x^2-y^2\right)\left(7z^2-r^2\right)$ & $\displaystyle \frac{\sqrt{5}}{2}xy\left(7z^2-r^2\right)$ \\
  & 3 & $\displaystyle \frac{\sqrt{70}}{4}zx\left(x^2-3y^2\right)$ & $\displaystyle \frac{\sqrt{70}}{4}yz\left(3x^2-y^2\right)$ \\
  & 4 & $\displaystyle \frac{\sqrt{35}}{8}\left(x^4-6x^2y^2+y^4\right)$ & $\displaystyle \frac{\sqrt{35}}{2}xy\left(x^2-y^2\right)$ \\ \hline
5 & 0 & $\displaystyle \frac{1}{8}\left(63z^5-70z^3r^2+15zr^4\right)$ & \\
  & 1 & $\displaystyle \frac{\sqrt{15}}{8}x\left[r^4+7z^2\left(3z^2-2r^2\right)\right]$ & $\displaystyle \frac{\sqrt{15}}{8}y\left[r^4+7z^2\left(3z^2-2r^2\right)\right]$ \\
  & 2 & $\displaystyle \frac{\sqrt{105}}{4}z\left(x^2-y^2\right)\left(3z^2-r^2\right)$ & $\displaystyle \frac{\sqrt{105}}{2}xyz\left(3z^2-r^2\right)$ \\
  & 3 & $\displaystyle \frac{\sqrt{70}}{16}x\left(x^2-3y^2\right)\left(9z^2-r^2\right)$ & $\displaystyle \frac{\sqrt{70}}{16}y\left(3x^2-y^2\right)\left(9z^2-r^2\right)$ \\
  & 4 & $\displaystyle \frac{3\sqrt{35}}{8}z\left(x^4-6x^2y^2+y^4\right)$ & $\displaystyle \frac{3\sqrt{35}}{2}xyz\left(x^2-y^2\right)$ \\
  & 5 & $\displaystyle \frac{3\sqrt{14}}{16}x\left(x^4-10x^2y^2+5y^4\right)$ & $\displaystyle \frac{3\sqrt{14}}{16}y\left(5x^4-10x^2y^2+y^4\right)$ \\ \hline
6 & 0 & $\displaystyle \frac{1}{16}\left(231z^6-315z^4r^2+105z^2r^4-5r^6\right)$ & \\
  & 1 & $\displaystyle \frac{\sqrt{21}}{8}zx\left[5r^4+3z^2\left(11z^2-10r^2\right)\right]$ & $\displaystyle \frac{\sqrt{21}}{8}yz\left[5r^4+3z^2\left(11z^2-10r^2\right)\right]$ \\
  & 2 & $\displaystyle \frac{\sqrt{210}}{32}\left(x^2-y^2\right)\left[r^4+3z^2\left(11z^2-6r^2\right)\right]$ & $\displaystyle \frac{\sqrt{210}}{16}xy\left[r^4+3z^2\left(11z^2-6r^2\right)\right]$\\
  & 3 & $\displaystyle \frac{\sqrt{210}}{16}zx\left(x^2-3y^2\right)\left(11z^2-3r^2\right)$ & $\displaystyle \frac{\sqrt{210}}{16}yz\left(3x^2-y^2\right)\left(11z^2-3r^2\right)$ \\
  & 4 & $\displaystyle \frac{3\sqrt{7}}{16}\left(x^4-6x^2y^2+y^4\right)\left(11z^2-r^2\right)$ & $\displaystyle \frac{3\sqrt{7}}{4}xy\left(x^2-y^2\right)\left(11z^2-r^2\right)$ \\
  & 5 & $\displaystyle \frac{3\sqrt{154}}{16}zx\left(x^4-10x^2y^2+5y^4\right)$ & $\displaystyle \frac{3\sqrt{154}}{16}yz\left(5x^4-10x^2y^2+y^4\right)$ \\
  & 6 & $\displaystyle \frac{\sqrt{462}}{32}\left[x^6-15x^2y^2\left(x^2-y^2\right)-y^6\right]$ & $\displaystyle \frac{\sqrt{462}}{16}xy\left(3x^4-10x^2y^2+3y^4\right)$ \\
\hline
\end{tabular}
\end{table*}

\begin{table*}[tb]
\caption{The cubic harmonics as linear combinations of the tesseral harmonics. The parity is given by $(-1)^p$.}
\label{cubic_sh}
\begin{minipage}{.45\textwidth}
\begin{tabular}{cccl}
\hline
$p$ & $\Gamma$ & $\gamma$ & $Z_{p,\Gamma,\gamma}=$ \\ \hline
1 & 4 & 1 & $\displaystyle Z_{11}^{\rm (c)}$ \\
  &   & 2 & $\displaystyle Z_{11}^{\rm (s)}$ \\
  &   & 3 & $\displaystyle Z_{10}$ \\ \hline
2 & 3 & 1 & $\displaystyle Z_{20}$ \\
  &   & 2 & $\displaystyle Z_{22}^{\rm (c)}$ \\
  & 5 & 1 & $\displaystyle Z_{21}^{\rm (s)}$ \\
  &   & 2 & $\displaystyle Z_{21}^{\rm (c)}$ \\
  &   & 3 & $\displaystyle Z_{22}^{\rm (s)}$ \\ \hline
3 & 2 & 1 & $\displaystyle Z_{32}^{\rm (s)}$ \\
  & 4 & 1 & $\displaystyle\frac{1}{2\sqrt{2}}\left(\sqrt{5}Z_{33}^{\rm (c)} - \sqrt{3}Z_{31}^{\rm (c)}\right)$ \\
  &   & 2 & $\displaystyle -\frac{1}{2\sqrt{2}}\left(\sqrt{5}Z_{33}^{\rm (s)} + \sqrt{3}Z_{31}^{\rm (s)}\right)$ \\
  &   & 3 & $\displaystyle Z_{30}$ \\
  & 5 & 1 & $\displaystyle -\frac{1}{2\sqrt{2}}\left(\sqrt{3}Z_{33}^{\rm (c)} + \sqrt{5}Z_{31}^{\rm (c)}\right)$ \\
  &   & 2 & $\displaystyle\frac{1}{2\sqrt{2}}\left(-\sqrt{3}Z_{33}^{\rm (s)} + \sqrt{5}Z_{31}^{\rm (s)}\right)$ \\
  &   & 3 & $\displaystyle Z_{32}^{\rm (c)}$ \\ \hline
4 & 1 & 1 & $\displaystyle\frac{1}{2\sqrt{3}}\left(\sqrt{5}Z_{44}^{\rm (c)} + \sqrt{7}Z_{40}\right)$ \\
  & 3 & 1 & $\displaystyle -\frac{1}{2\sqrt{3}}(\sqrt{7}Z_{44}^{\rm (c)} - \sqrt{5}Z_{40})$ \\
  &   & 2 & $\displaystyle -Z_{42}^{\rm (c)}$ \\
  & 4 & 1 & $\displaystyle -\frac{1}{2\sqrt{2}}\left(Z_{43}^{\rm (s)} + \sqrt{7}Z_{41}^{\rm (s)}\right)$ \\
  &   & 2 & $\displaystyle -\frac{1}{2\sqrt{2}}\left(Z_{43}^{\rm (c)} - \sqrt{7}Z_{41}^{\rm (c)}\right)$ \\
  &   & 3 & $\displaystyle Z_{44}^{\rm (s)}$ \\
  & 5 & 1 & $\displaystyle \frac{1}{2\sqrt{2}}\left(\sqrt{7}Z_{43}^{\rm (s)} - Z_{41}^{\rm (s)}\right)$ \\
  &   & 2 & $\displaystyle -\frac{1}{2\sqrt{2}}\left(\sqrt{7}Z_{43}^{\rm (c)} + Z_{41}^{\rm (c)}\right)$ \\
  &   & 3 & $\displaystyle Z_{42}^{\rm (s)}$ \\ \hline
\end{tabular}
\end{minipage}
\begin{minipage}{.45\textwidth}
\begin{tabular}{cccl}
\hline
$p$ & $\Gamma$ & $\gamma$ & $Z_{p,\Gamma,\gamma}=$ \\ \hline
5 & 3 & 1 & $\displaystyle Z_{54}^{\rm (s)}$ \\
  &   & 2 & $\displaystyle -Z_{52}^{\rm (s)}$ \\
  & 4a & 1 & $\displaystyle \frac{1}{8\sqrt{2}}\left(3\sqrt{7}Z_{55}^{\rm (c)} - \sqrt{35}Z_{53}^{\rm (c)} + \sqrt{30}Z_{51}^{\rm (c)}\right)$ \\
  &    & 2 & $\displaystyle \frac{1}{8\sqrt{2}}\left(3\sqrt{7}Z_{55}^{\rm (s)} + \sqrt{35}Z_{53}^{\rm (s)} + \sqrt{30}Z_{51}^{\rm (s)}\right)$ \\
  &    & 3 & $\displaystyle Z_{50}$ \\
  & 4b & 1 & $\displaystyle \frac{1}{16}\left(\sqrt{10}Z_{55}^{\rm (c)} + 9\sqrt{2}Z_{53}^{\rm (c)} + 2\sqrt{21}Z_{51}^{\rm (c)}\right)$ \\
  &    & 2 & $\displaystyle \frac{1}{16}\left(\sqrt{10}Z_{55}^{\rm (s)} - 9\sqrt{2}Z_{53}^{\rm (s)} + 2\sqrt{21}Z_{51}^{\rm (s)}\right)$ \\
  &    & 3 & $\displaystyle Z_{54}^{\rm (c)}$ \\
  & 5 & 1 & $\displaystyle \frac{1}{4\sqrt{2}}\left(-\sqrt{15}Z_{55}^{\rm (c)} - \sqrt{3}Z_{53}^{\rm (c)} + \sqrt{14}Z_{51}^{\rm (c)}\right)$ \\
  &   & 2 & $\displaystyle \frac{1}{4\sqrt{2}}\left(\sqrt{15}Z_{55}^{\rm (s)} - \sqrt{3}Z_{53}^{\rm (s)} - \sqrt{14}Z_{51}^{\rm (s)}\right)$ \\
  &   & 3 & $\displaystyle Z_{52}^{\rm (c)}$ \\ \hline
6 & 1 & 1 & $\displaystyle \frac{1}{2\sqrt{2}}\left(-\sqrt{7}Z_{64}^{\rm (c)} + Z_{60}\right)$ \\
  & 2 & 1 & $\displaystyle \frac{1}{4}\left(-\sqrt{5}Z_{66}^{\rm (c)} + \sqrt{11}Z_{62}^{\rm (c)}\right)$ \\
  & 3 & 1 & $\displaystyle \frac{1}{2\sqrt{2}}\left(Z_{64}^{\rm (c)} + \sqrt{7}Z_{60}\right)$ \\
  &   & 2 & $\displaystyle \frac{1}{4}\left(\sqrt{11}Z_{66}^{\rm (c)} + \sqrt{5}Z_{62}^{\rm (c)}\right)$ \\
  & 4 & 1 & $\displaystyle \frac{1}{8}\left(-\sqrt{22}Z_{65}^{\rm (s)} - \sqrt{30}Z_{63}^{\rm (s)} + 2\sqrt{3}Z_{61}^{\rm (s)}\right)$ \\
  &   & 2 & $\displaystyle \frac{1}{8}\left(\sqrt{22}Z_{65}^{\rm (c)} - \sqrt{30}Z_{63}^{\rm (c)} - 2\sqrt{3}Z_{61}^{\rm (c)}\right)$ \\
  &   & 3 & $Z_{64}^{\rm (s)}$ \\
  & 5a & 1 & $\displaystyle \frac{1}{16}\left(\sqrt{3}Z_{65}^{\rm (s)} + \sqrt{55}Z_{63}^{\rm (s)} + 3\sqrt{22}Z_{61}^{\rm (s)}\right)$ \\
  &    & 2 & $\displaystyle \frac{1}{16}\left(\sqrt{3}Z_{65}^{\rm (c)} - \sqrt{55}Z_{63}^{\rm (c)} + 3\sqrt{22}Z_{61}^{\rm (c)}\right)$ \\
  &    & 3 & $Z_{66}^{\rm (s)}$ \\
  & 5b & 1 & $\displaystyle \frac{1}{16}\left(\sqrt{165}Z_{65}^{\rm (s)} - 9Z_{63}^{\rm (s)} + \sqrt{10}Z_{61}^{\rm (s)}\right)$ \\
  &    & 2 & $\displaystyle \frac{1}{16}\left(\sqrt{165}Z_{65}^{\rm (c)} + 9Z_{63}^{\rm (c)} + \sqrt{10}Z_{61}^{\rm (c)}\right)$ \\
  &    & 3 & $\displaystyle Z_{62}^{\rm (s)}$ \\
\hline
\end{tabular}
\end{minipage}
\end{table*}

\begin{table*}[tb]
\caption{The generalized Stevens' factors for the Hund's-rule ground multiplet in the Russell-Saunders scheme.}
\label{stevens}
\begin{tabular}{cccccccc}
\hline
& Ce$^{3+}$ & Pr$^{3+}$ & Nd$^{3+}$ & Pm$^{3+}$ & Sm$^{3+}$ & Eu$^{3+}$ & Gd$^{3+}$ \\ \hline
$g_n^{(0)}=n$ & 1 & 2 & 3 & 4 & 5 & 6 & 7 \\
$J$ & $5/2$ & 4 & $9/2$ & 4 & $5/2$ & 0 & $7/2$ \\
$L$ & 3 & 5 & 6 & 6 & 5 & 3 & 0 \\
$S$ & $1/2$ & 1 & $3/2$ & 2 & $5/2$ & 3 & $7/2$ \\
\hline
$g_n^{(2)}$ & $\frac{-2}{5\cdot 7}$ & $\frac{-2^2\cdot13}{3^2\cdot5^2\cdot11}$ & $\frac{-7}{3^2\cdot11^2}$ &   $\frac{2\cdot7}{3\cdot5\cdot11^2}$ & $\frac{13}{3^2\cdot5\cdot7}$ & 0 & 0  \\
$g_n^{(4)}$ & $\frac{2}{3^2\cdot5\cdot7}$ & $\frac{-2^2}{3^2\cdot5\cdot11^2}$ & $\frac{-2^3\cdot17}{3^3\cdot11^3\cdot13}$ & $\frac{2^3\cdot7\cdot17}{3^3\cdot5\cdot11^3\cdot13}$ & $\frac{2\cdot13}{3^3\cdot5\cdot7\cdot11}$ & 0 & 0  \\
$g_n^{(6)}$ & 0 & $\frac{2^4\cdot17}{3^4\cdot5\cdot7\cdot11^2\cdot13}$ & $\frac{-5\cdot17\cdot19}{3^3\cdot7\cdot11^3\cdot13^2}$ & $\frac{2^3\cdot17\cdot19}{3^3\cdot7\cdot11^3\cdot13^2}$ & 0 & 0 & 0 \\
\hline
$g_n^{(1)}$ & $\frac{2\cdot3}{7}$ & $\frac{2^2}{5}$ & $\frac{2^3}{11}$ & $\frac{3}{5}$ & $\frac{2}{7}$ & 0 & 2 \\
$g_n^{(3)}$ & $\frac{-2}{5\cdot7}$ & $\frac{2\cdot13}{3^2\cdot5^2\cdot11}$ & $\frac{2\cdot5\cdot7}{3\cdot11^2\cdot13}$ & $\frac{7}{5\cdot11^2}$ & $\frac{-2\cdot13}{3^3\cdot5\cdot7}$ & $0$ & 0 \\
$g_n^{(5)}$ & $\frac{2^2}{3^2\cdot7\cdot11}$ & $\frac{-2^2}{3^2\cdot7\cdot11^2}$ & $\frac{2^3\cdot5\cdot17}{3^4\cdot11^3\cdot13}$ & $\frac{2^2\cdot17}{3^3\cdot11^3\cdot13}$ & $\frac{-2^2\cdot13}{3^2\cdot7\cdot11^2}$ & $0$ & 0 \\
\hline
$r_n^{(1)}$ & $-4$ & $-3$ & $-7/3$ & $-7/4$ & $-6/5$ & -- & 0 \\
$r_n^{(3)}$ & $-5/2$ & 0 & 4 & $-4$ & 0 & -- & 0 \\
$r_n^{(5)}$ & $-2$ & $-2$ & $3/2$ & $-3/2$ & 2 & -- & 0 \\
\hline
& Tb$^{3+}$ & Dy$^{3+}$ & Ho$^{3+}$ & Er$^{3+}$ & Tm$^{3+}$ & Yb$^{3+}$ \\ \hline
$g_n^{(0)}=n$ & 8 & 9 & 10 & 11 & 12 & 13 \\
$J$ & 6 & $15/2$ & 8 & $15/2$ & 6 & $7/2$ \\
$L$ & 3 & 5 & 6 & 6 & 5 & 3 \\
$S$ & 3 & $5/2$ & 2 & $3/2$ & 1 & $1/2$ \\
\hline
$g_n^{(2)}$ & $\frac{-1}{3^2\cdot11}$ & $\frac{-2}{3^2\cdot5\cdot7}$ & $\frac{-1}{2\cdot3^2\cdot5^2}$ & $\frac{2^2}{3^2\cdot5^2\cdot7}$ & $\frac{1}{3^2\cdot11}$ & $\frac{2}{3^2\cdot7}$  \\
$g_n^{(4)}$ & $\frac{2}{3^3\cdot5\cdot11^2}$ & $\frac{-2^3}{3^3\cdot5\cdot7\cdot11\cdot13}$ & $\frac{-1}{2\cdot3\cdot5\cdot7\cdot11\cdot13}$ & $\frac{2}{3^2\cdot5\cdot7\cdot11\cdot13}$ & $\frac{2^3}{3^4\cdot5\cdot11^2}$ & $\frac{-2}{3\cdot5\cdot7\cdot11}$ \\
$g_n^{(6)}$ & $\frac{-1}{3^4\cdot7\cdot11^2\cdot13}$ & $\frac{2^2}{3^3\cdot7\cdot11^2\cdot13^2}$ & $\frac{-5}{3^3\cdot7\cdot11^2\cdot13^2}$ & $\frac{2^3}{3^3\cdot7\cdot11^2\cdot13^2}$ & $\frac{-5}{3^4\cdot7\cdot11^2\cdot13}$ & $\frac{2^2}{3^3\cdot7\cdot11\cdot13}$ \\
\hline
$g_n^{(1)}$ & $\frac{3}{2}$ & $\frac{2^2}{3}$ & $\frac{5}{2^2}$ & $\frac{2\cdot3}{5}$ & $\frac{7}{2\cdot3}$ & $\frac{2^3}{7}$ \\
$g_n^{(3)}$ & $\frac{-7}{2^2\cdot3\cdot5\cdot11}$ & $\frac{-2^2}{3^3\cdot7\cdot13}$ & $\frac{1}{2^2\cdot5^3}$ & $\frac{2^2\cdot61}{3\cdot5^2\cdot7\cdot11\cdot13}$ & $\frac{1}{2\cdot3^2\cdot5\cdot11}$ & $\frac{-2^4}{5\cdot7\cdot13}$ \\
$g_n^{(5)}$ & $\frac{13}{2^3\cdot3^4\cdot11^2}$ & $\frac{-2^2\cdot47}{3^5\cdot7\cdot11^2\cdot13}$ & $\frac{1}{3^3\cdot5\cdot7\cdot11\cdot13}$ & $\frac{2^2\cdot31}{3^3\cdot7\cdot11^3\cdot13}$ & $\frac{-7}{2\cdot3^5\cdot11^2}$ & $\frac{2^3\cdot5}{3^3\cdot7\cdot11\cdot13}$ \\
\hline
$r_n^{(1)}$ & $1/2$ & 1 & $3/2$ & 2 & $5/2$ & 3 \\
$r_n^{(3)}$ & $4/3$ & 0 & $-25/4$ & $55/6$ & 0 & $-13$ \\
$r_n^{(5)}$ & $4/9$ & $27/20$ & $-5/3$ & $22/9$ & $-9/2$ & $-13/3$ \\
\hline
\end{tabular}
\end{table*}


\begin{thebibliography}{99}
\bibitem{xortho} For $\Gamma_{4u}$ symmetry, $\hat{J}_{p,4u,\gamma}$'s are not orthogonal to others. Thus, we have mixed two different $\hat{J}_{p,\Gamma,\gamma}$ operators in accordance with the Schmitt's re-orthonormalization.
\bibitem{Kubo04a} The expression looks different from that of [\citen{Kubo04}], but the term with $\overline{M}_{3,5u}$ is equivalent to that of [\citen{Kubo04}].
\end{thebibliography}
\end{document}